\documentclass[useAMS,usenatbib]{mn2e_mod}
\usepackage{deluxetable}
\usepackage[utf8]{inputenc}
\usepackage{amsmath}
\usepackage{amsfonts}
\usepackage{amssymb}
\usepackage{graphicx}
\usepackage[font=small,labelfont=bf]{caption}
\usepackage[a4paper]{geometry}
\usepackage{tablefootnote}
\usepackage{natbib}
\title {Photometric and Spectroscopic Properties of Type II-P Supernovae}

		\author[Faran et al.]
		{T. Faran$^{1}$, 
		D. Poznanski$^{1}$\thanks{dovi@tau.ac.il},
		A. V. Filippenko$^{2}$,
		R. Chornock$^{3}$,
		R. J. Foley$^{4,5}$,\newauthor
		M. Ganeshalingam$^{2,6}$,		
		D. C. Leonard$^{7}$,
		W. Li$^{2,8}$,
		M. Modjaz$^{9}$, 
		E. Nakar$^{1}$,\newauthor
		F. J. D. Serduke$^{2}$,
		and J. M. Silverman$^{10,11}$\\
		\\
		$^{1}$School of Physics and Astronomy, Tel-Aviv University, Tel Aviv 69978, Israel.\\
		$^{2}$Department of Astronomy, University of California, Berkeley, CA 94720-3411, USA.\\
		$^{3}$Harvard-Smithsonian Center for Astrophysics, 60 Garden Street, Cambridge, MA 02138, USA.\\
		$^{4}$Astronomy Department, University of Illinois at Urbana-Champaign, 1002 W.\ Green Street, Urbana, IL 61801 USA\\
		$^{5}$Department of Physics, University of Illinois Urbana-Champaign, 1110 W.\ Green Street, Urbana, IL 61801 USA\\
		$^{6}$Lawrence Berkeley National Laboratory, Berkeley, CA 94720, USA.\\
		$^{7}$Department of Astronomy, San Diego State University, San Diego, CA 92182, USA.\\
        $^{8}$Deceased 12 December 2011.\\
		$^{9}$CCPP, New York University, 4 Washington Place New York City, NY, 10003, USA.\\
		$^{10}$Department of Astronomy, University of Texas at Austin, Austin, TX 78712, USA.\\
		$^{11}$NSF Astronomy and Astrophysics Postdoctoral Fellow.\\
		}

\begin{document}
	\maketitle
	\label{firstpage}
	\begin{abstract}
We study a sample of 23 Type II Plateau supernovae (SNe~II-P), all observed with the same set of instruments. Analysis of their  photometric evolution confirms that their typical plateau duration is 100 days with little scatter, showing a tendency to get shorter for more energetic SNe. The rise time from explosion to plateau does not seem to correlate with luminosity. We analyze their spectra, measuring typical ejecta velocities, and confirm that they follow a well behaved power-law decline. We find indications of high-velocity material in the spectra of six of our SNe. We test different dust extinction correction methods by asking the following -- does the uniformity of the sample increase after the application of a given method? A reasonably behaved underlying distribution should become tighter after correction. No method we tested made a significant improvement. 

\end{abstract}

\section{Introduction}

Stars more massive than $\sim 8$~M$_{\odot}$ end their lives as core-collapse supernovae (CC~SNe). This is based on multiple lines of evidence, from their locations, in or near regions of star formation, to multiple progenitor identifications (see \citealt{Smartt:2009} for a review). The explosions of stars that have kept a large hydrogen envelope until their demise are observationally defined as Type II SNe, based on the presence of broad hydrogen lines in their spectra with line widths of several thousands km s$^{-1}$. The lines typically have P-Cygni profiles, formed by the expanding hydrogen-rich ejecta. 

Type II SNe are a diverse class, with a large range of observed luminosities, light-curve shapes, and spectroscopic features. These have been historically used to further subclassify these events. The optical light-curve shape has been used to separate  those that decline linearly with time (II-L) from those that show a pronounced plateau (II-P; \citealt{Barbon:1979}). Later, two more classes were defined spectroscopically, SNe~IIn that have relatively narrow hydrogen emission lines attributed to interaction of the ejecta with circumstellar matter, and SNe~IIb that are spectroscopically intermediate between SNe~II-P and SNe~Ib (see \citealt{Filippenko:1997} for a review). SNe~Ib are interpreted as arising from stars stripped of their envelope, as indicated by their helium dominated, hydrogen-free spectra. SNe~IIb early-time spectra evolve from SN~II-P-like hydrogen-rich spectra, to SN~Ib-like \citep{Filippenko:1988}. 

Individual SNe~II-P have been well studied in the past thanks to their proximity. Examples include SN\,1999em (e.g., \citealt{Baron:2000}; \citealt{Leonard:2002a}; \citealt{Dessart:2006}), and SN\,1999gi \citep{Leonard:2002b}. Both objects have progenitor mass estimates -- a very tight restriction on the upper mass of the SN\,1999em progenitor was derived by \citet{Smartt:2009b}, who found 15~M$_{\odot}$, and for SN\,1999gi \citet{Smartt:2001} report an upper limit of 9$^{+3}_{-2}$~M$_{\odot}$. Another example of a nearby, well-studied event is SN\,2005cs (e.g., \citealt{Pastorello:2009}); it had a low luminosity and a low-mass progenitor. 

However, there are very few studies of samples of SNe~II-P. \citet{Hamuy:2002} and \citet{Hamuy:2003b} analyzed a sample of $\sim 20$ SNe~II-P, and found relations between plateau luminosity, ejecta velocities, and nickel abundances, as well as a useful luminosity-velocity relation which allows for distance measurements. This is an empirical application of the expanding photosphere method (EPM; \citealt{Kirshner:1974}; \citealt{Eastman:1996}; \citealt{Dessart:2005b}; \citealt{Dessart:2008}; \citealt{Jones:2009}). This method and its potential for cosmological distance measurements was further studied by \citet{Nugent:2006}, \citet{Poznanski:2009}, \citet{DAndrea:2010}, \citet{Poznanski:2010}, and \citet{Olivares:2010}. \citet{Maguire:2010b} examined its extension into the infrared. \citet{Arcavi:2012} analyzed a small sample of SNe~II-P, contrasting them with other SNe~II, and found that while they differ wildly in luminosity, they seem to have similar plateau durations, all near 100 days. 

While this paper was in final editing stages \citet{Anderson:2014} published an analysis of Type II SN photometry. They present a sample of more than 100 V-band light curves. Their results are consistent with some of our photometric findings, namely that these SNe seem to form a continuous class, with a range of luminosities, with a correlation between plateau duration and brightness. However, they consider a larger range of SNe, including objects that we would classify as II-L and analyze separately in a companion paper (Faran et al. 2014). 

Here we present a sample of 23 SNe~II-P, many of which have not been studied before, all observed by the Lick Observatory SN Search (LOSS; \citealt{Filippenko:2001}). All objects have high-cadence photometry in 3--4 bands, and between few and tens of optical spectra. This results in a dataset with 1574 photometric points and 152 spectra. We further use this sample to examine methods for dust-extinction correction\footnote{This paper is dedicated to the memory of our dear friend and colleague, Dr. Weidong Li, without whom these data would not exist.}.

\section{Observations}

\begin{table*}
\caption{SN Sample}
%\tablewidth{0pt}
%\tabletypesize{\scriptsize}
\begin{tabular}{l c c c c}
\hline\hline
SN name &
 $z_{\rm host}$&
$\mu$\tablenotemark{a}&
Explosion MJD&
$E(B-V)_{\rm MW}$\tablenotemark{c}\\

\hline
1999bg &  0.0043 & 31.83  & 51251 (14) & 0.0184 \\
1999D  &  0.0104 & 33.17\tablenotemark{b}  & -& 0.0166 \\
1999em &  0.0024 & 29.95  & 51476 (4) & 0.0400\\
1999gi &  0.0020 & 29.55  & 51518 (4) & 0.0166\\
2000bs &  0.0280 & 35.30\tablenotemark{b}  & 51648 (6) & 0.0317 \\
2000dj &  0.0154 & 34.19  & 51788 (7) & 0.0734 \\
2001bq &  0.0087 & 32.78\tablenotemark{b}  & 52034 (6) & 0.0441 \\
2001cm &  0.0114 & 33.48  & 52063 (1) & 0.0124 \\
2001hg &  0.0086 & 33.00  & - & 0.0351  \\
2001X  &  0.0049 & 31.37  & 51963 (5) & 0.0401 \\
2002an &  0.0129 & 33.85  & 52288 (8) & 0.0348  \\
2002bx &  0.0075 & 34.42 & 52354 (10) & 0.0118 \\
2002ca &  0.0109 & 33.26\tablenotemark{b}  & 52353 (15) & 0.0242 \\
2002gd &  0.0089 & 33.07  & 52552 (?) & 0.0676 \\
2002hh &  0.0001 & 28.87  & 52576 (3) & 0.3413 \\
2003gd &  0.0022 & 29.66  & 52735 (67) & 0.0682 \\
2003hl &  0.0082 & 32.68  &52867 (5) & 0.0734 \\
2003iq &  0.0082 & 32.68  & 52920 (2) & 0.0725  \\
2003Z  &  0.0043 & 31.77  & 52664 (5) & 0.0384 \\
2004du &  0.0168 & 34.40  & 53228 (2) & 0.0947 \\
2004et &  0.0001 & 28.87  & 53271 (1) & 0.3415\\
2005ay &  0.0027 & 31.15  & 53442 (14) & 0.0214 \\
2005cs &  0.0015 & 29.51  & 53545 (4) & 0.0347 \\
\hline
\end{tabular}
\tablenotetext{a}{Distance modulus (mag) from NED, unless noted otherwise.}
\tablenotetext{b}{Redshift-based distance.}
\tablenotetext{c}{Mag, assuming $R_V = 3.1$.}
\label{t:SNedataP1}
\end{table*}

Imaging was obtained with the 0.76-m Katzman Automatic Imaging Telescope (KAIT), the SN search engine of LOSS. The sample was compiled from events for which precise photometry could be obtained -- that is, for which template images were obtained after the SN had faded. Since the focus on the study is SNe~II-P, we only include objects with a pronounced plateau in the light curve, removing SNe~II-L, IIb, IIn, and peculiar SNe, such as SN\,2000cb, a SN\,1987A-like object \citep{Kleiser:2011}. The distinction between II-P and II-L events will be defined by Faran et al. (2014b).

KAIT images of the SNe were reduced as described by \citet{Poznanski:2009}, \citet{Ganesh:2010}, and \citet{Silverman:2012}. Briefly, after standard image reduction, image subtraction was performed, using galaxy templates obtained with KAIT on photometric nights, at least a few months after the SN had faded beyond detection. Photometry was obtained using differential point-spread function fitting using the DAOPHOT package in IRAF, in order to measure the SN flux relative to local standards in the field. Calibrations were obtained on photometric nights using both the KAIT and the 1~m Nickel telescope at Lick Observatory. We correct the magnitudes for Galactic extinction using the maps of \citet{Schlegel:1998}.

\begin{figure*}
\centering
\includegraphics[width=1\textwidth]{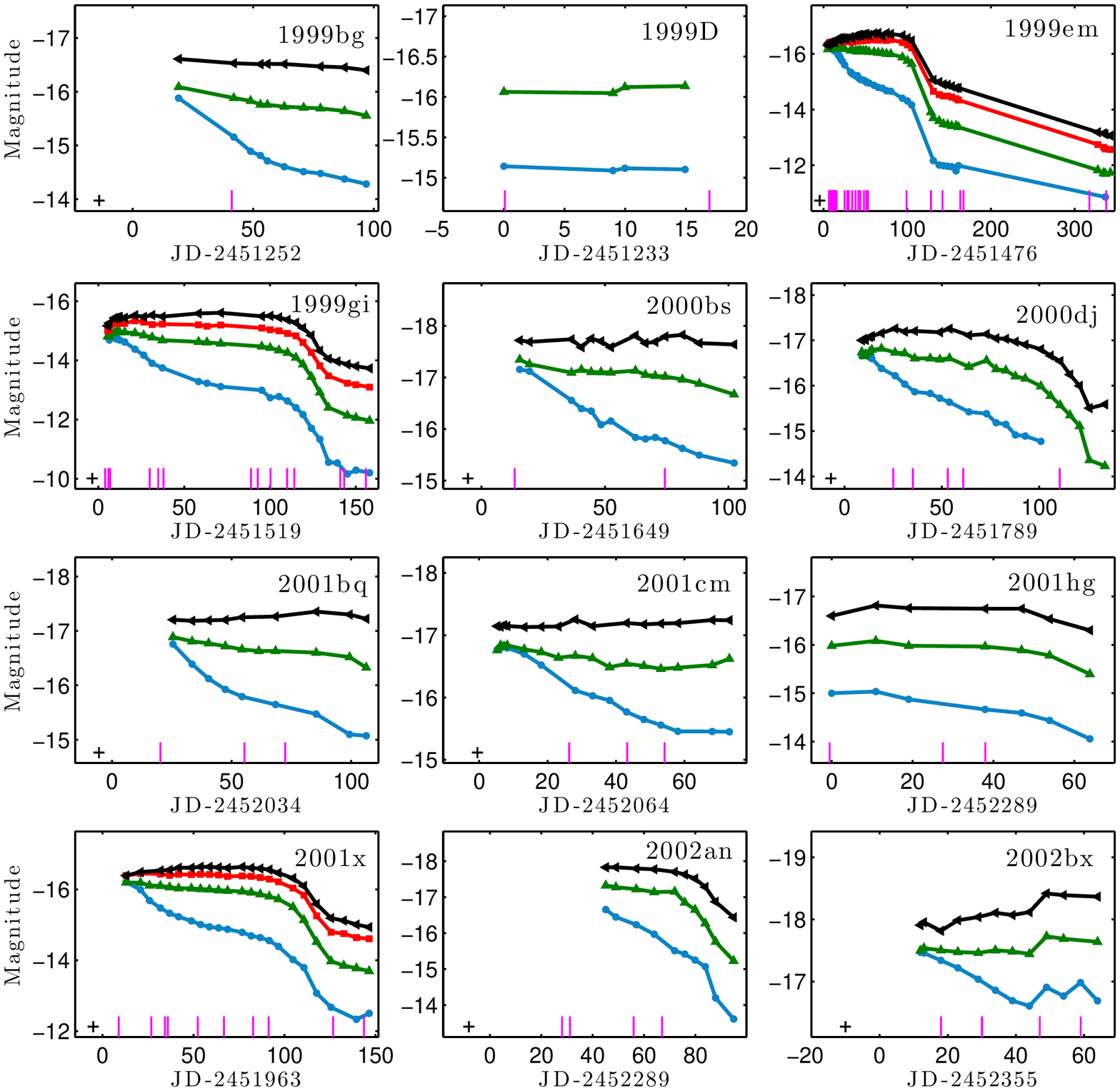}
\caption{Light curves of the 23~SNe in our sample. $B$ (blue circles), $V$ (green triangles), $R$ (red squares), and $I$ (black side triangles). Magenta ticks mark the epochs at which spectra were obtained. The black crosses are the last nondetections. SNe\,1999D and 2001hg are plotted relative to the first photometric point. Continued in Figure \ref{f:phot_grid2}}.\label{f:phot_grid}
\end{figure*}

\begin{figure*}
 \centering
 \includegraphics[width=1\textwidth]{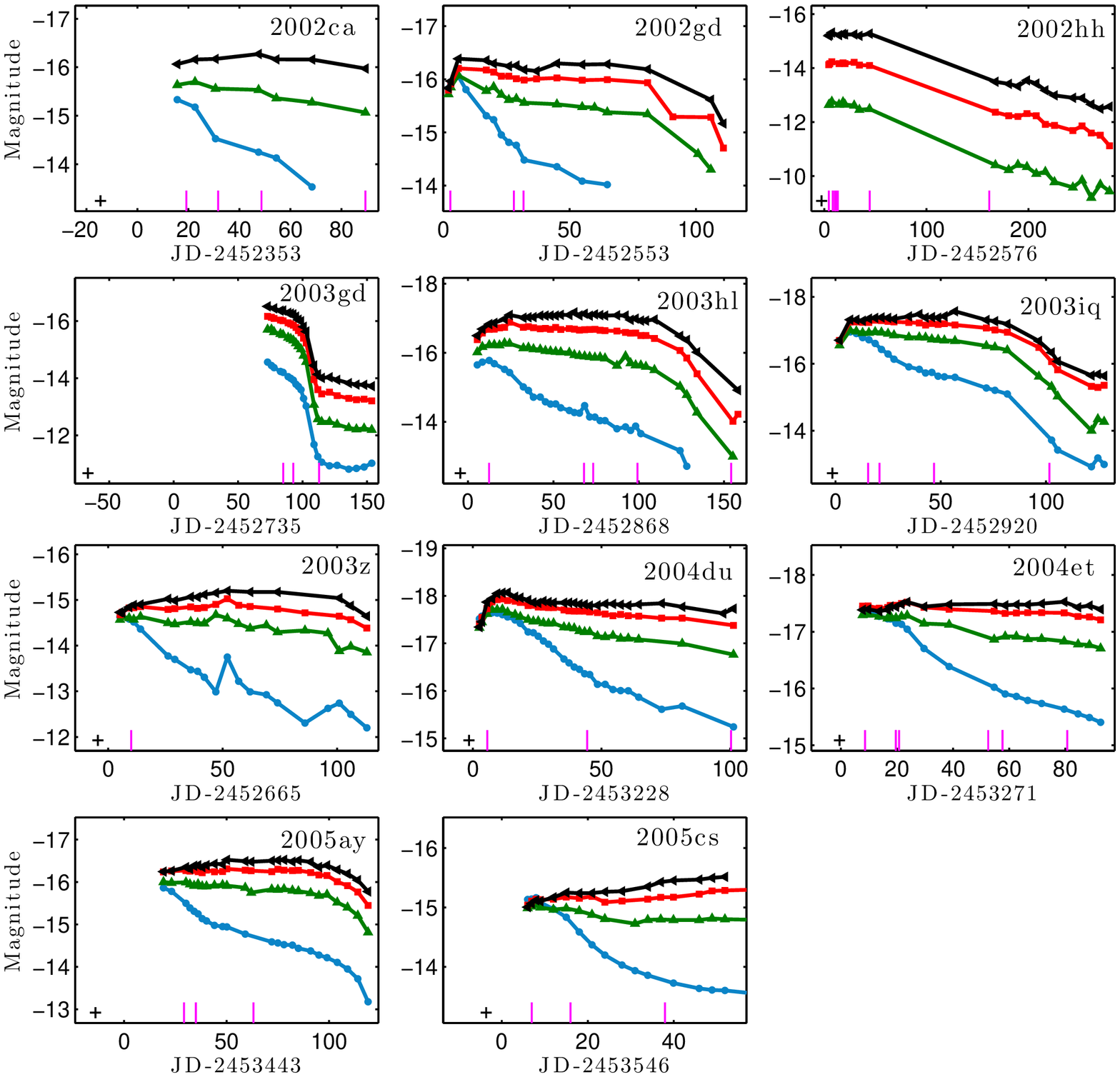}
\caption{Light curves of the 23~SNe in our sample, continued. }\label{f:phot_grid2}
 \end{figure*}

In Figures \ref{f:phot_grid} and \ref{f:phot_grid2} we show the $BVRI$ light curves of our 23~SNe, marking the epochs at which spectra were taken. 
The light curves are typically better sampled than the timescales for brightness change, so it is safe to interpolate the photometry when needed. Information on the SNe is presented in Table \ref{t:SNedataP1}. The explosion day is set as the midpoint between the first detection and the last nondetection, and the uncertainty is conservatively set as half the difference. There are three objects with weak constraints. For SN\,2002gd a rough estimate for the explosion day can be obtained by comparing the epoch of peak brightness to SNe with similar light curves. The photometry of SN\,1999D and SN\,2001hg does not even allow that. These objects are naturally excluded from analyses where time reference is required.

Distance measurements are collected from NED\footnote{The NASA/IPAC Extragalactic Database (NED) is operated by the Jet Propulsion Laboratory, California Institute of Technology, under contract with the National Aeronautics and Space Administration (NASA).} and averaged, using only distances based on the Tully-Fisher method, Cepheids, and SNe~Ia. When the distance is ambiguous or unknown, we derive the distance to the SN using Hubble's law and the redshift, using a Hubble constant of H$_0 = 73$~km~s$^{-1}$~Mpc$^{-1}$, assuming an uncertainty of 10\%. All of the objects in our sample are at low redshifts with $z < 0.03$. 

We obtained optical spectra with the Kast double spectrograph (Miller \& Stone 1993) mounted on the Lick Observatory 3-m Shane telescope, the Low Resolution Imaging Spectrometer (LRIS; \citealt{Oke:1995}) on the 10-m Keck-I telescope, the Stover spectrograph mounted on the 1-m Nickel telescope and the Deep Imaging Multi-Object Spectrograph (DEIMOS; \citealt{Faber:2003}) on the 10-m Keck-II telescope. Reduction details are given by \citet{Poznanski:2009} and \citet{Li:2001}.

Partial photometric and spectroscopic data are shown in Tables \ref{t:phot} - \ref{t:spec}. Complete data are available in the electronic version.

All photometric and spectroscopic data are available electronically via the Berkeley\footnote{http://hercules.berkeley.edu/database/} \citep{Silverman:2012}  and WISeREP\footnote{http://www.weizmann.ac.il/astrophysics/wiserep/} \citep{Yaron:2012} databases.

\section{Extinction Estimation}
\label{s: extinction}

There are different common methods for extinction and reddening correction. However, various methods often give different results (see, e.g., Table 4 of \citealt{Olivares:2010}). In this section, we inclusively try all the techniques we can apply in search of an optimal correction.

\citet{Schmidt:1992} showed that the $B-V$ curves of different SNe~II-P evolve similarly, while some objects appear to be offset by a constant. Those SNe were found in highly inclined spiral galaxies and consequently suffered from more severe host extinction than the remainder, which reside in face-on spiral galaxies or in the outer regions of their host galaxies. The shift in colour can be attributed to the combined effect of dust and intrinsic differences. 

The standard deviation in colour of a sample probes a combination of intrinsic diversity and redenning. If all objects are intrinsically similar, and are behind dust with a similar composition, then $\sigma_{B-V} \approx E(B-V)$, the typical colour excess. For our sample, when using epochs between days 5 and 150 after explosion, and excluding objects with explosion day uncertainties larger than 10 days, we find a value $\sigma_{B-V} = 0.22$~mag. This is within the range of typically measured values of $E(B-V)$ (see, e.g., \citealt{Maguire:2010}, Table 3; \citealt{Pastorello:2003}, Table 1, and references therein).

\subsection{Increasing uniformity}
\label{s:Eliminating Intrinsic Diversity}

A-priori, there could be a phase in the light curve of a SN where the prevailing physical conditions are similar, so that scatter in colour may be attributed mostly to extinction. At the beginning of the plateau phase, the entire hydrogen envelope is ionised, and consequently highly opaque owing to electron scattering. As the temperature in the outer layers of the ejecta approaches 6000~K, the plateau phase begins -- hydrogen starts to recombine and the opacity drops by several orders of magnitude. A recombination front develops toward the inner layers, until it reaches the base of the hydrogen envelope. At this stage the internal energy has been depleted, and the plateau ends (e.g., \citealt{Popov:1993,Kasen:2009}). However, variations of the photospheric temperature could result from variations in H/He abundance ratio \citep{Arnett:1996}, which alter the hydrogen density in the envelope and hence shift the recombination temperature.

Nevertheless, \citet{Hamuy:2005} suggest that during the plateau, the colour scatter should be minimal since the same hydrogen recombination temperature is reached. They derive the colour excess for several objects by comparing colour curves at the end of the plateau to a template SN (the well-studied SN\,1999em), assuming that all SNe reach the same temperature at that phase.

We align the colour curves in time and measure $\sigma_{B-V}$, where we try two different time origins -- the explosion day and the end of the plateau. We define the latter as the day by which the brightness drops by 0.5~mag below the average plateau luminosity. 

\begin{figure}
\centering
\includegraphics[width=1\columnwidth]{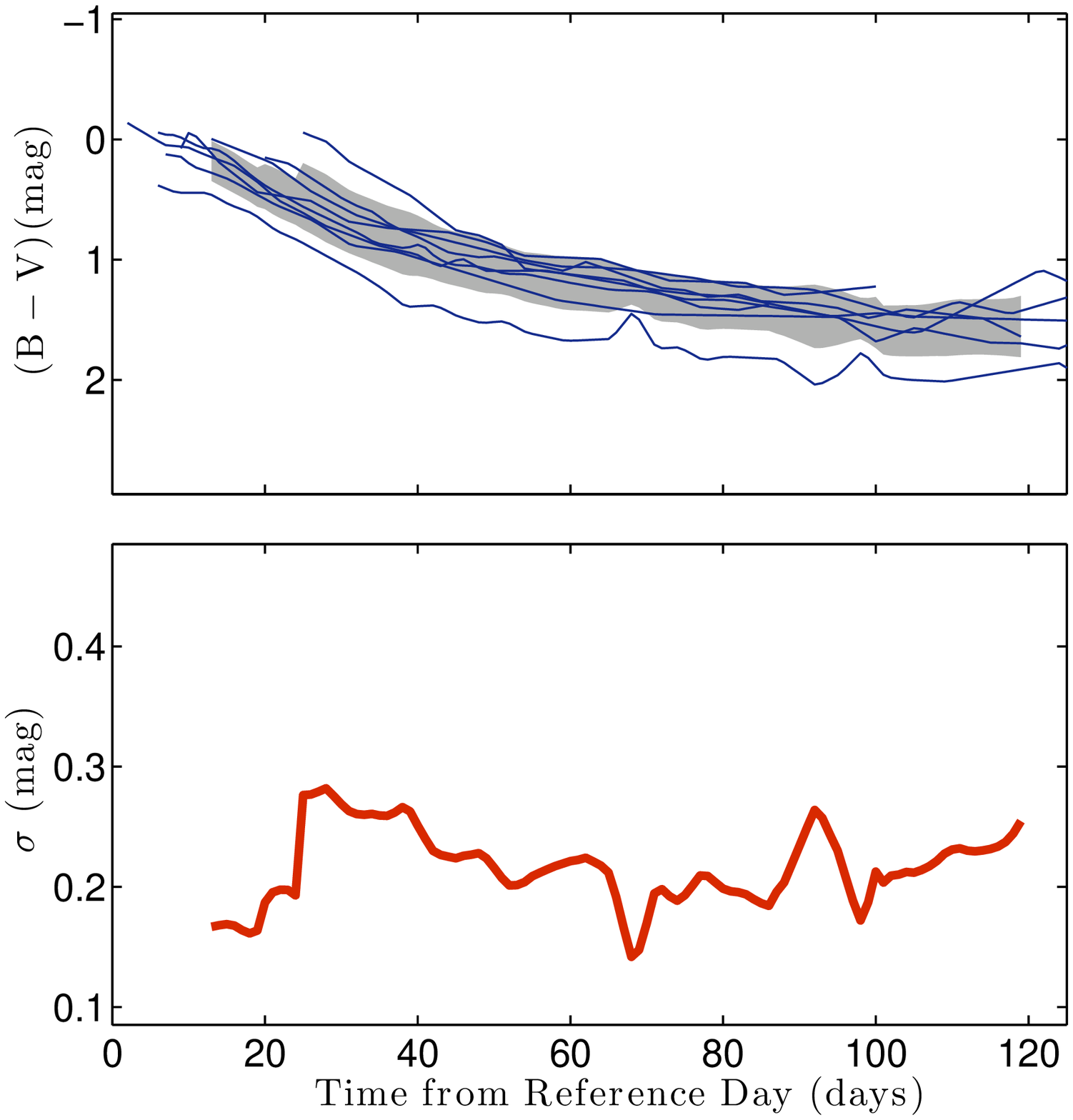} 
\caption{$B-V$ colour, as a function of time (upper panel) and its standard deviation (bottom panel). Colour curves are aligned according to their explosion day. The shaded area represents $\pm 1\sigma$  from the mean colour values. The scatter appears similar throughout the plateau.}
\label{f: color_std_texp}

\includegraphics[width=1\columnwidth]{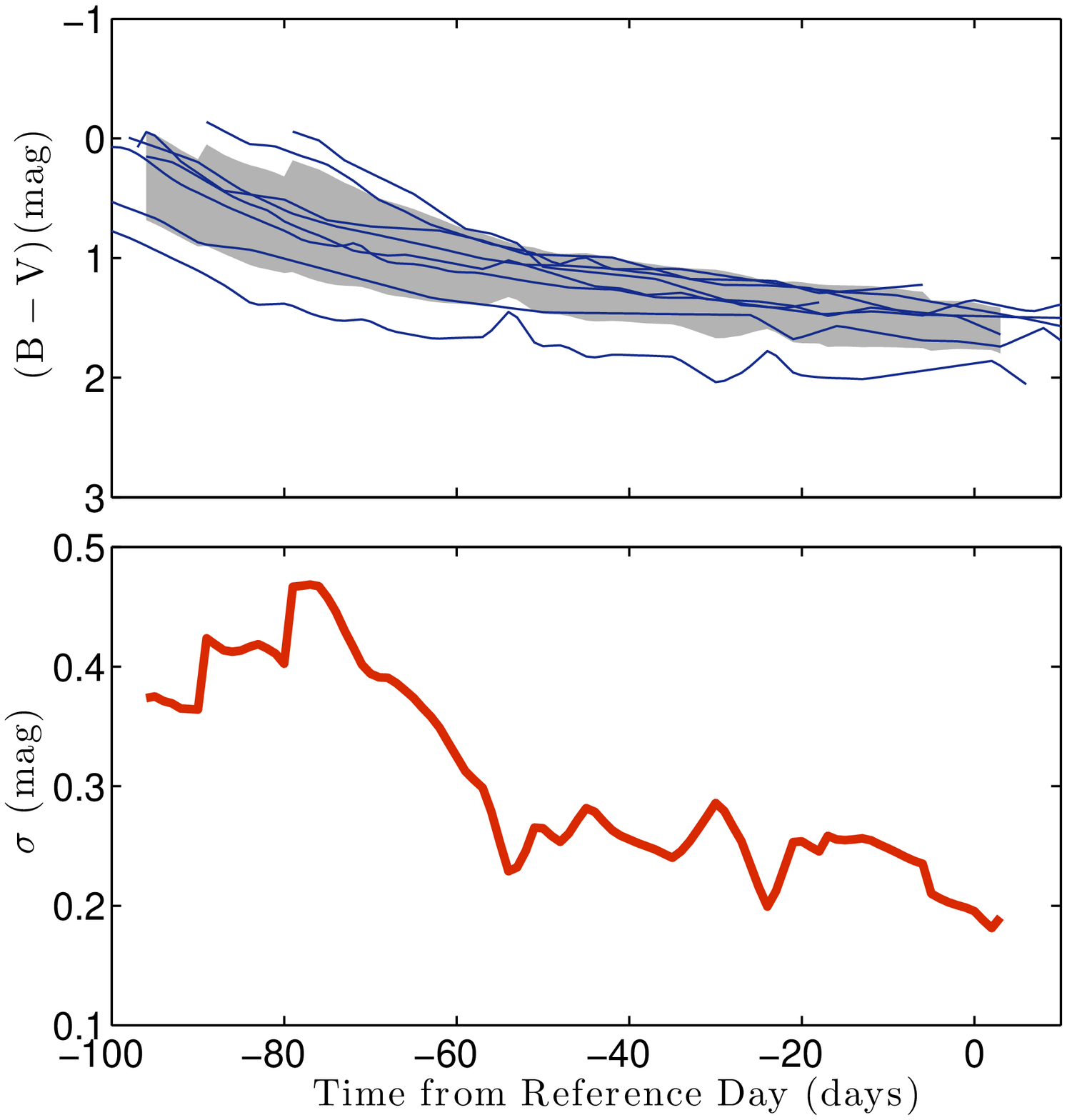} 
\centering
\caption{Same as Figure \ref{f: color_std_texp}, with curves aligned at the end of the plateau phase. The curves are more discrepant than in Figure \ref{f: color_std_texp} where the explosion date was taken as origin.}
\label{f: color_std_tp}
\end{figure}

We plot the time evolution of the $B-V$ curves aligned to the explosion day (Figure \ref{f: color_std_texp}) and to the plateau end, $t_{p}$ (Figure \ref{f: color_std_tp}), along with their corresponding $\sigma_{B-V}$ on the bottom panels. Note that we were able to extract $t_{p}$ for only a fraction of the SNe (9 out of 23~SNe). For consistency, we use the same subsample for both analyses according to the two time origins.

When the curves are aligned to the explosion day, the standard deviation remains generally constant over time, showing that there is no ``preferred epoch'' with smaller scatter. When we set the zero time to the end of the plateau phase (e.g., as done by \citealt{Olivares:2010}), $\sigma_{B-V}$ at that origin is similar to what we had before, but rises dramatically as we move away toward earlier times. This indicates that not only intrinsic differences still exist at all plateau phases, but that the end of the plateau is not a natural origin for object comparison. This is consistent with a theoretical picture, where the precise end of the plateau will depend on envelope masses, velocities, nickel mass, and other specific properties of a given object \citep[e.g., ][]{Kasen:2009}. 

Nevertheless, we follow \citet{Hamuy:2005}, take the end of the plateau as a reference point, and compare colours at that point to those of SN\,1999em. We assume that any relative departure of the colour of a SN from SN\,1999em is caused by dust. This SN has been extensively modeled by \citet{Baron:2000} and others and has an upper limit of $E(B-V)_{\rm tot} = 0.15$~mag. This value also includes Milky Way reddening in the direction of SN\,1999em (0.036~mag;  \citealt{Schlegel:1998}), so we adopt $E(B-V)_{\rm host} = 0.12$~mag. The results are shown in the last column of Table \ref{t: E(B-V)}. 

\begin{figure}
\centering
\includegraphics[width=1\columnwidth]{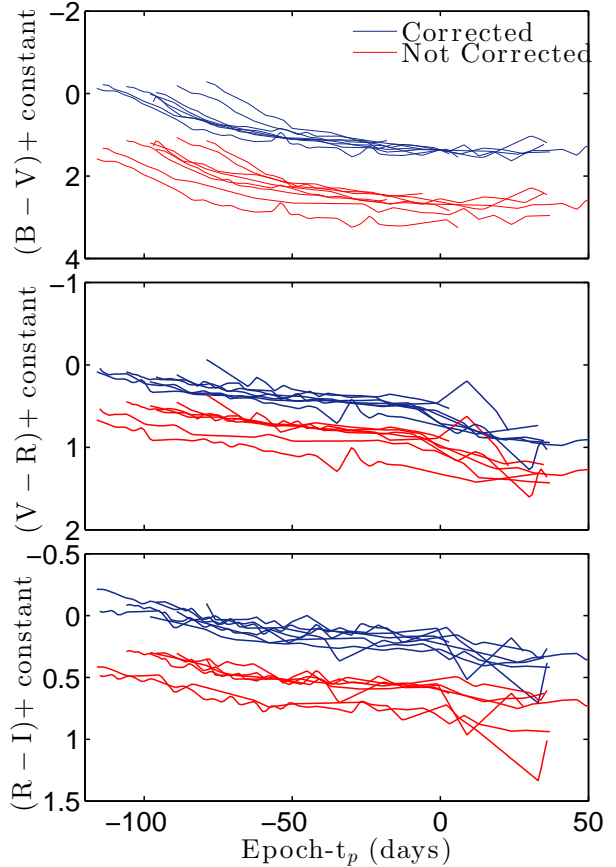} 
\caption{
$B-V$, $V-R$, and $R-I$ curves aligned to the end of the plateau phase, before and after extinction correction by comparing to SN\,1999em. Since $B-V$ colour is used for correction its scatter is obviously reduced (though not significantly at early times), but other bands hardly benefit from this correction.}
\label{f: colors_compare}
\end{figure}

Using the resulting $E(B-V)$ values and the extinction curve of \citet{Cardelli:1989}, we can estimate $E(V-R)$ and $E(R-I)$. In Figure \ref{f: colors_compare} we show the curves before and after the correction. The corrected $B-V$ curves obviously meet at zero, and show a gradual decline in scatter approaching the end of the plateau. This sample is small and does not allow for a definitive conclusion. However, one can see that while the scatter in the redder colours is marginally reduced, it is achieved by shifting two objects, at the cost of somewhat scattering the remainder of the sample. The same can be seen in $V-I$, where colour is not affected by the evolution of the H$\alpha$ line.  Since before any correction the scatter in colour is small, only highly accurate extinction values can have a chance to reduce it more.

\subsection{The Na~I~D equivalent width}
Early studies on SNe~Ia have suggested that the equivalent width of the Na~I~D doublet, EW$_{\rm Na~I~D}$, is a good proxy for the amount of extinction toward a SN (\citealt{Barbon:1990}; \citealt{Turatto:2003}). However, using a much larger sample, \citet{Poznanski:2011} showed that although EW$_{\rm Na~I~D}$ correlates with the extinction when measured from low-resolution spectra, there is  so much scatter that the method is effectively useless. This is likely due to a combination of the doublet lines not being resolved, variations in observing conditions introducing different amounts of continuum light from the host galaxy, the effects of circumstellar matter \citep{Phillips:2013}, as well as intrinsic sources of scatter, such as variations in dust-to-gas ratios in different galaxy types (\citealt{Issa:1990}; \citealt{Lisenfeld:1998}), the effects of depletion of metals on dust grains (e.g., Savage \& Mathis 1979), and similar processes. 

We collect from the literature four of the available dereddening relations: \citet{Barbon:1990}, $E(B-V) = 0.25\,{\rm EW}_{\rm Na~I~D}$ (Bar025); the two linear relations derived by \citet{Turatto:2003}, $E(B-V) = 0.16\,{\rm EW}_{\rm Na~I~D} - 0.01$ (Tur016) and $E(B-V) =-0.04 + 0.51\,{\rm EW}_{\rm Na~I~D}$ (Tur051); and the relation derived by \citet{Poznanski:2011}: $E(B-V) = -0.08 + 0.43\,{\rm EW}_{\rm Na~I~D}$ (Poz043). We note that we do not use the more recent relation from \citet{Poznanski:2012}, since it only holds for EW $< 1$~\AA, and many objects in the sample exceed this value.

The EWs are measured using a similar method to the one described by \citet{Poznanski:2011}. We inspect the spectra visually, and when we detect a line, we identify its edges and pseudo-continuum manually. We fit a low-order polynomial to the continuum and divide the data by the continuum fit. We then perform a numerical integral on the normalised data to find the EW and measure the noise of the spectrum, $N$, defined as the 5$\sigma$ clipped standard deviation around the flattened spectrum. We assume the uncertainty in EW to be dominated by $N$. We calculate the uncertainty, $\delta$EW, as $\delta {\rm EW} = N\,W/2$, where $W$ is the width of the line (defined as the distance in \AA\ between the manually marked line limits). We divide by a factor of 2 to account for the approximately triangular shape of the feature. We average the EWs obtained from different spectra for every SN, except for SN\,1999em for which we use only one spectrum, where the doublet lines were resolved, and we assume this particular spectrum to be more reliable than the rest of the spectra. We manage to detect Na~I~D in a total of 11~SNe, and their values appear in Table \ref{t: E(B-V)}.

\begin{figure}
\centering
\includegraphics[width=1\columnwidth]{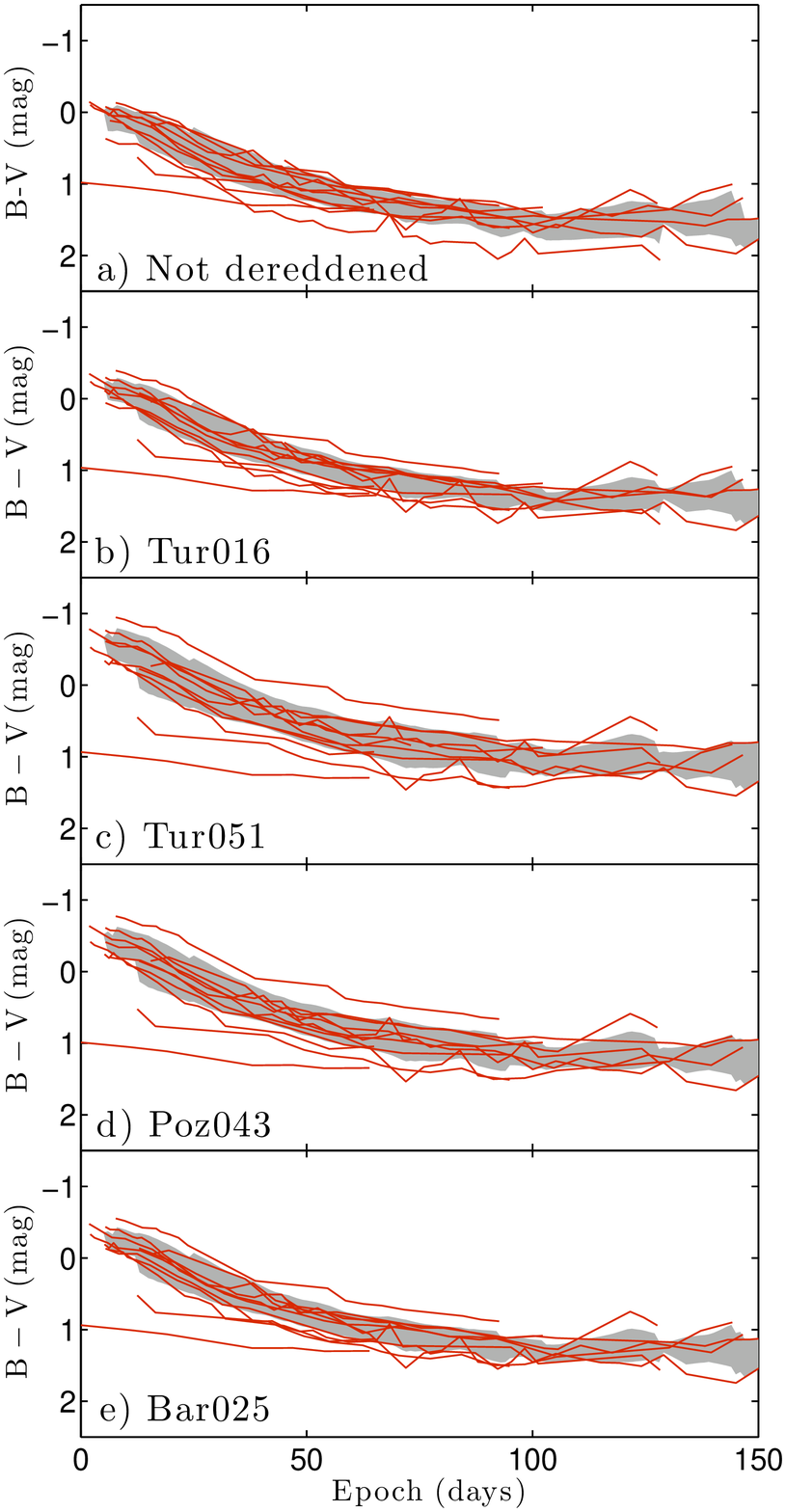} 
\caption{$B-V$ colour curves of 12~SNe for which a Na~I~D absorption line is detected, dereddened according to various relations from the literature; a decrease in scatter is observed only in (b), where the average $\sigma$ is reduced by 9\%, whereas (c) and (d) show an increase of 40\% and 15\% (respectively) in scatter. The shaded area represents $\pm 1 \sigma$ from the mean colour. }
\label{f: NeEW dereddened scatter}
\end{figure}

From the dereddened colour curves shown in Figure \ref{f: NeEW dereddened scatter}, it is clear that Tur051 and Poz043 only increase the scatter and therefore overestimate $E(B-V)$. Tur016 reduces $\sigma$ only by an insignificant 9\% and Bar025 by 7\%. 

Comparison of $E(B-V)$ values extracted from the three methods that showed a modicum of success can be found in Table \ref{t: E(B-V)}. The methods give different and typically inconsistent values, indicating that they are too often equivalent to an educated guess. Since all the methods seem to suffer from systematic uncertainties, any averaging between the results is nearly guaranteed to be wrong. With the exception of SN\,2002hh \citep{Pozzo:2006}, our SNe suffer rather moderate extinction, and we apply no dust correction in the following sections.

\begin{table*}
\caption{Extinction values extracted from the Na~I~D EW}
\begin{tabular}{l c c c c}
\hline\hline
SN name &
EW$_{\rm Na~I~D}$ (\AA) &
$E(B-V)_{\rm Tur016}$ &
$E(B-V)_{\rm Bar025}$&
$E(B-V)_{\rm color}$\\
\hline

1999em & 1.43& $\rm 0.22 \pm0.04$ & $\rm 0.36 \pm0.06$  & 0.12  \\ 
1999gi &0.92& $\rm 0.14 \pm0.01$ & $\rm 0.23 \pm0.02$  & 0.33 \\
2000bs & 0.97& $\rm 0.15 \pm0.04$ & $\rm 0.24 \pm0.07$  & - \\
2000dj & -&  - & -  & -0.09  \\ 
2001cm & 0.73& $\rm 0.11 \pm0.03$ & $\rm 0.18 \pm0.04$  & - \\
2001X & 0.51&$\rm 0.07 \pm0.01$ & $\rm 0.13 \pm0.01$  & 0.09  \\
2002an & 0.43& $\rm 0.06 \pm0.01$ & $\rm 0.11 \pm0.01$  & - \\
2002gd & 0.92& $\rm 0.14 \pm0.01$ & $\rm 0.23 \pm0.02$  & 0.27 \\
2002hh & 2.75& $\rm 0.43 \pm0.02$ & $\rm 0.69 \pm0.03$  & - \\
2003hl & 2.00& $\rm 0.31 \pm0.03$ & $\rm 0.50 \pm0.05$ & 0.57 \\
2003iq & 1.33& $\rm 0.20 \pm0.02$ & $\rm 0.33 \pm0.03$  & -0.03 \\
2003Z & - & - & - & 0.19 \\
2004du &- & - & -  & 0.13 \\
2004et & 1.67& $\rm 0.26 \pm0.00$ & $\rm 0.42 \pm0.01$  & -  \\
2005ay & - & - & -  & 0.11 \\

\hline
\end{tabular}
\label{t: E(B-V)}
\end{table*}

\section{Photometric properties}\label{s:photometry}

\citet{Pastorello:2004} suggest that there are two distinct classes of SNe~II-P, faint and bright. \citet{Fraser:2011} further find that low-luminosity objects also tend to result from explosions of lower-mass stars. In contrast, using all SNe~II-P with measured progenitor masses, \citet{Poznanski:2013} find that there is a correlation between the mass and ejecta velocity, and since velocity and luminosity are related as well, mass should correlate with luminosity. With the large uncertainties involved in mass determination, it is hard to determine whether what one sees is indeed a correlation in a one-parameter family or two distinct populations. 

\begin{figure}
\centering
\includegraphics[width=1\columnwidth]{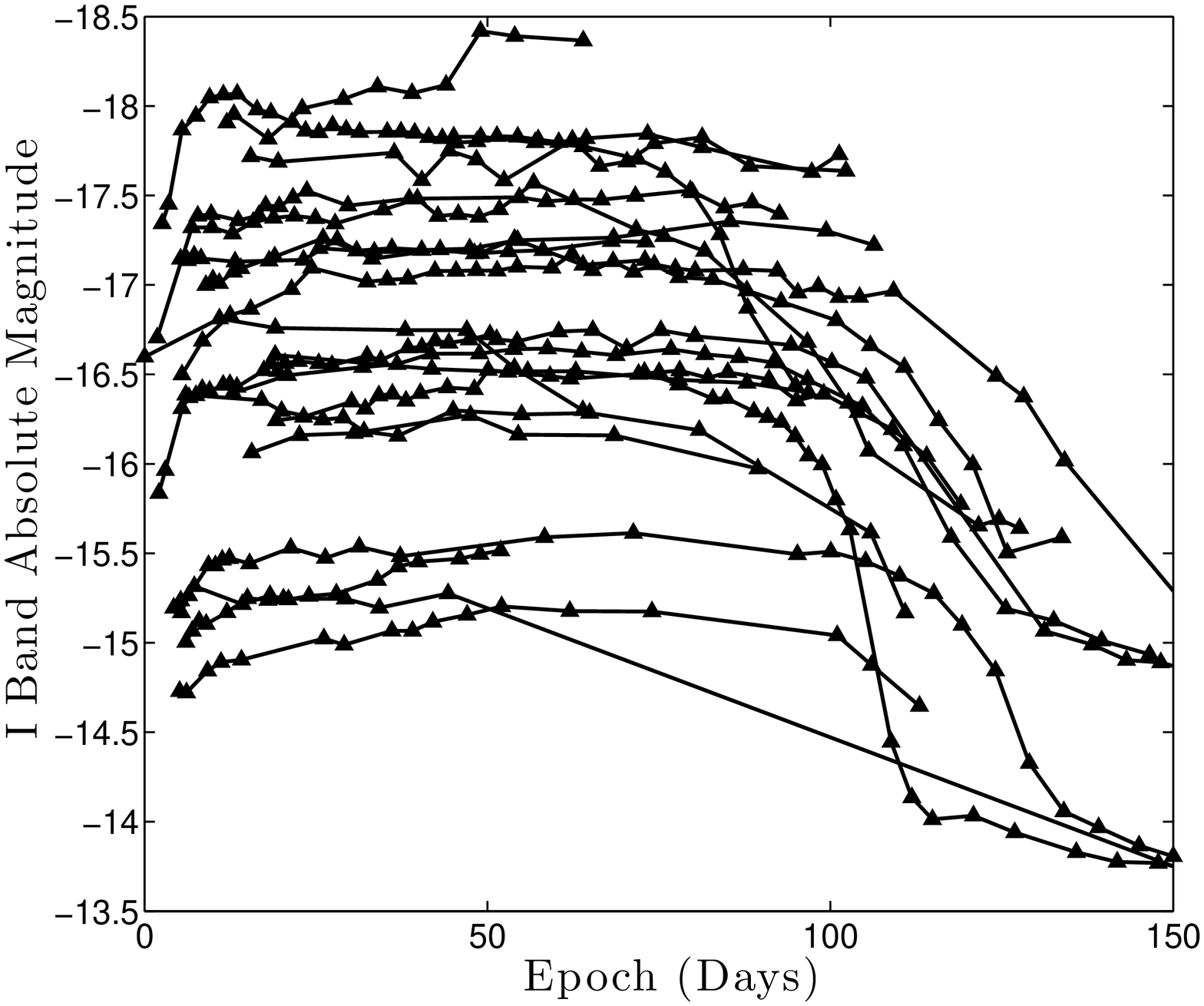}
\caption{$I$-band light curves in absolute magnitude for the sample. The curves seem to continuously cover a range of order 3~mag (between $-18$ and $-15$ mag). However, plateau duration is rather uniform, near 100 days. }
\label{f:abs_mag}
\end{figure}

In Figure \ref{f:abs_mag} we show the $I$-band light curves for the entire sample. It is rather apparent that while it spans a range of 3~mag, there is a continuous distribution of luminosities, with perhaps a half-magnitude gap below $M=-16$\,mag.

\begin{figure}
\centering
\includegraphics[width=1\columnwidth]{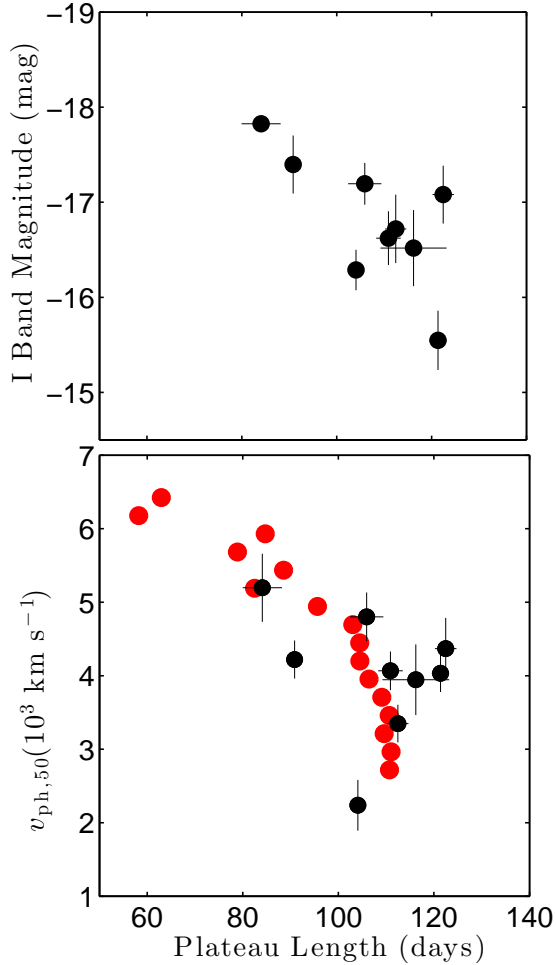}

\caption{Top panel: $I$-band absolute magnitude on day 50 vs. plateau duration. The plateau length stays approximately constant over a large range of  magnitudes, though perhaps getting shorter for brighter SNe. Bottom panel: photospheric velocities on day 50 vs. plateau duration. Since there is an established correlation between luminosity and velocity for SNe~II-P, the plot resembles the one in the top panel. The red dots are a subset of grid models computed by \citet{Dessart:2010}, interpolated to satisfy the mass-velocity relation found by \citet{Poznanski:2013}. These models show very good agreement with our measurements.}
\label{f:M_v_plateaulength}
\end{figure}

In the top panel of Figure \ref{f:M_v_plateaulength} we plot the $I$-band absolute magnitude on day 50 vs. plateau length, defined to be the day where the light curve drops by 0.5~mag from the mean plateau value. We exclude objects with incomplete data (i.e., for which the light curves terminate before the post-plateau drop), so that we are ultimately left with 9 objects. The analysis is done in the $I$ band where the plateau is the most prominent. Plateau durations are rather similar for all objects, of order 100 days, reinforcing the findings of \citet{Arcavi:2012} and \citet{Poznanski:2013}. However, within the scatter, it is apparent that more-luminous SNe have shorter plateaus, as predicted by \citet{Poznanski:2013}.

The luminosity of SNe~II-P correlates with the day 50 photospheric velocity \citep{Hamuy:2002}, and in the lower panel of Figure \ref{f:M_v_plateaulength} we plot the latter against the plateau length. We compare our results to a subset of grid models computed by \citet{Dessart:2010}, interpolated to satisfy the mass-velocity relation found by \citet{Poznanski:2013} (red points). We can see a reasonable match between the models and our observations of SNe~II-P, both consistent with a uniform plateau duration of $\sim 100$ days up to a velocity $\sim 4500$~km~s$^{-1}$, after which the plateaus tend to get shorter.

\begin{figure}
\centering
\includegraphics[width=1\columnwidth]{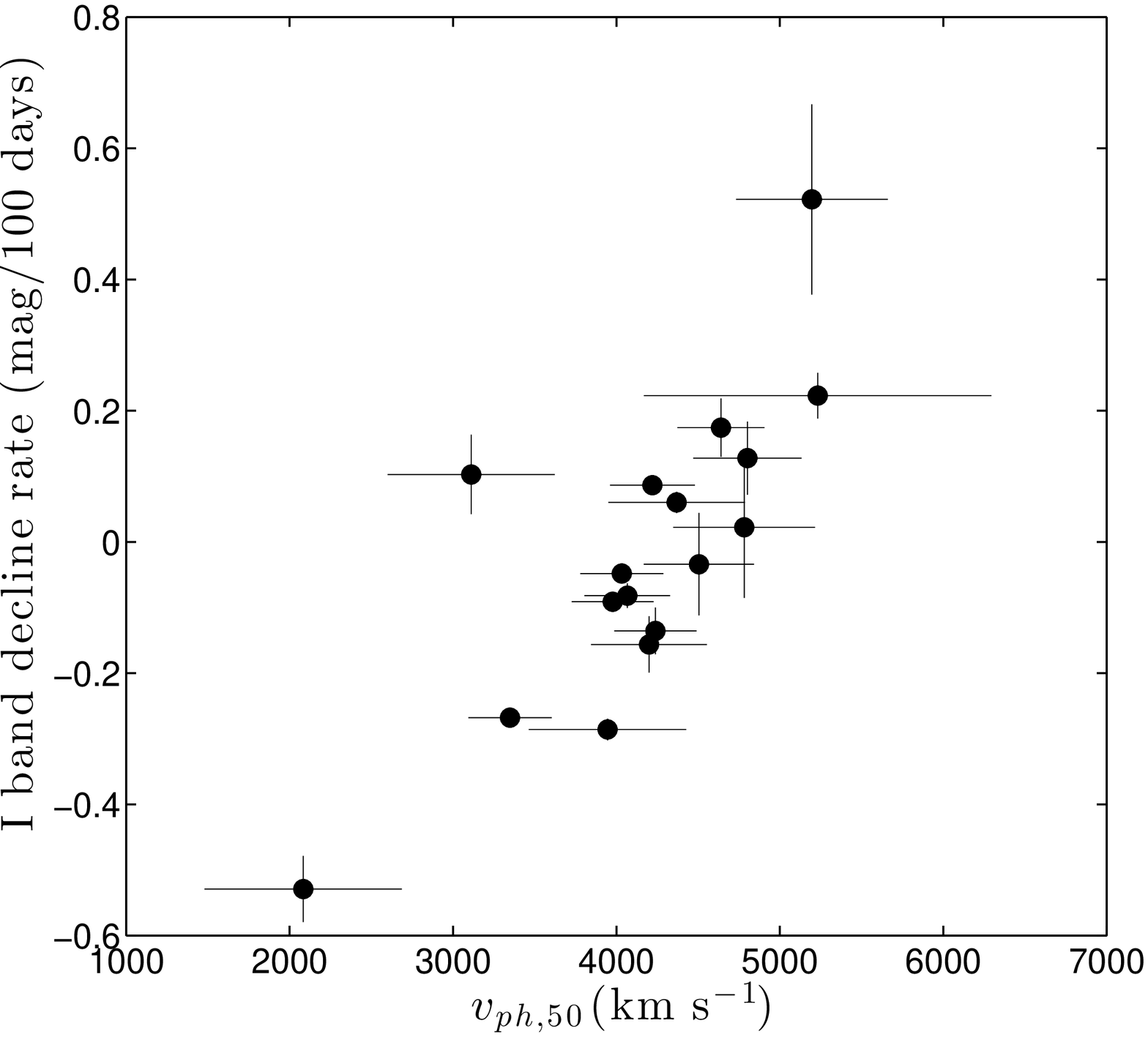}

\caption{The $I$-band decline rate vs. $v_{\rm ph,50}$. There is a clear correlation between the decline rate of the light curve and the photospheric velocity.}
\label{f:decrate_v50}
\end{figure}

We also find that the decline rate of the $I$-band light curves correlates well with $v_{\rm ph,50}$, as seen in Figure \ref{f:decrate_v50}. A possible explanation is that high ejecta velocities will cause a fast decrease in density, which will be followed by a quicker release of radiation.

\subsection{Colour evolution}
\begin{figure}
\centering
\includegraphics[width=1\columnwidth]{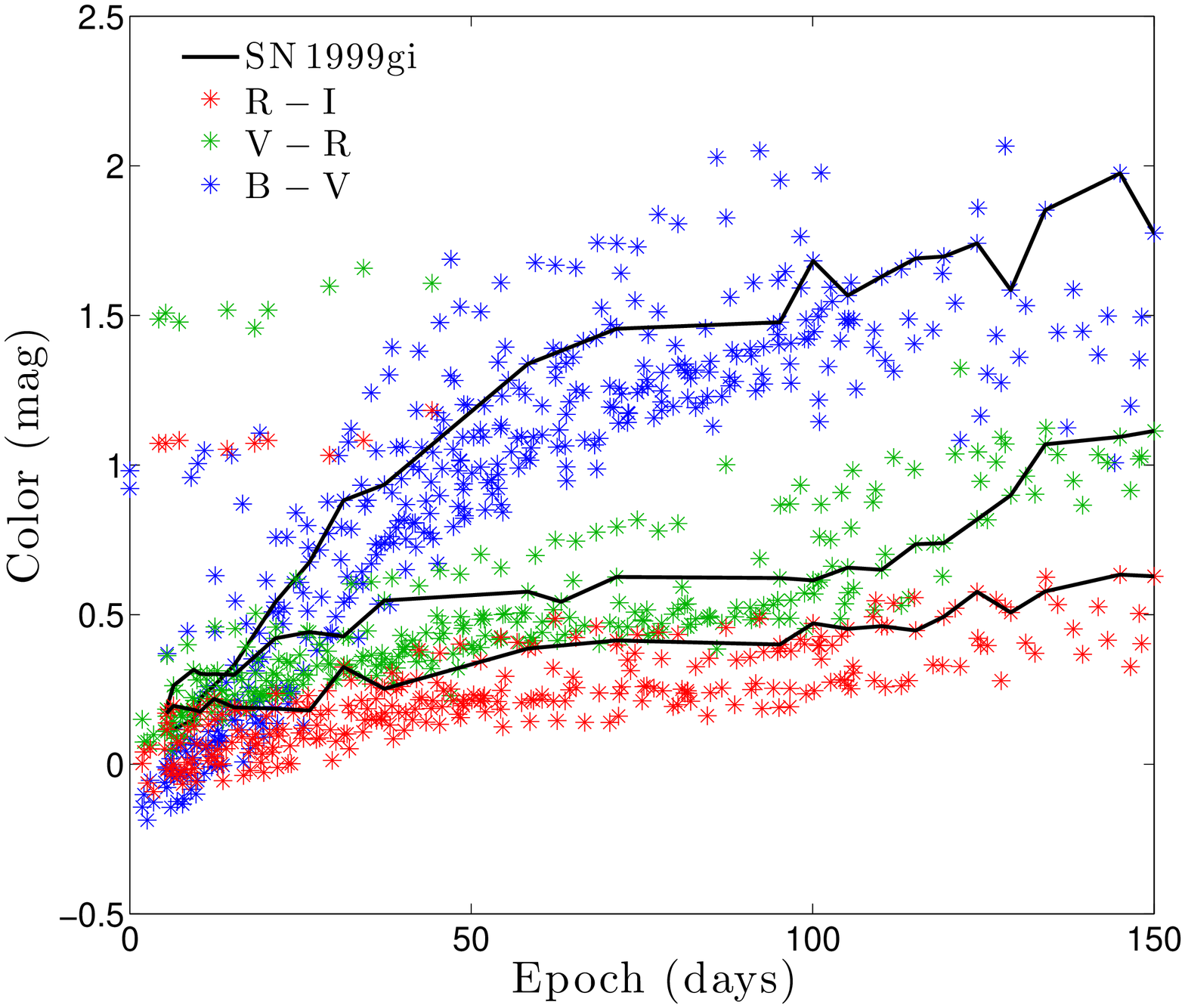}
\caption{$B-V$, $V-R$, and $V-I$ colour evolution of the whole sample. The individual curves of SN\,1999gi are shown in black to guide the eye. The highly reddened object separated from the other curves is SN\,2002hh.}
\label{f:color_curves}
\end{figure}

Figure \ref{f:color_curves} shows the colour evolution of the sample. At early times, the $B-V$ colour declines sharply owing to the drop in temperature. This is because the bluer bands are in the Wien part of the blackbody spectrum and therefore more sensitive to temperature change. In addition, strong iron blanketing mostly affects bluer bands. However, during the recombination phase $B-V$ changes moderately since the conditions at the photosphere remain approximately constant. For the same reason, $V-R$ and $R-I$ show almost no change. The larger scatter in $B-V$ compared to redder colours could be due to extinction, but also to variations in metallicities that translate to a different amount of line blanketing.

\begin{figure}
\centering
\includegraphics[width=1\columnwidth]{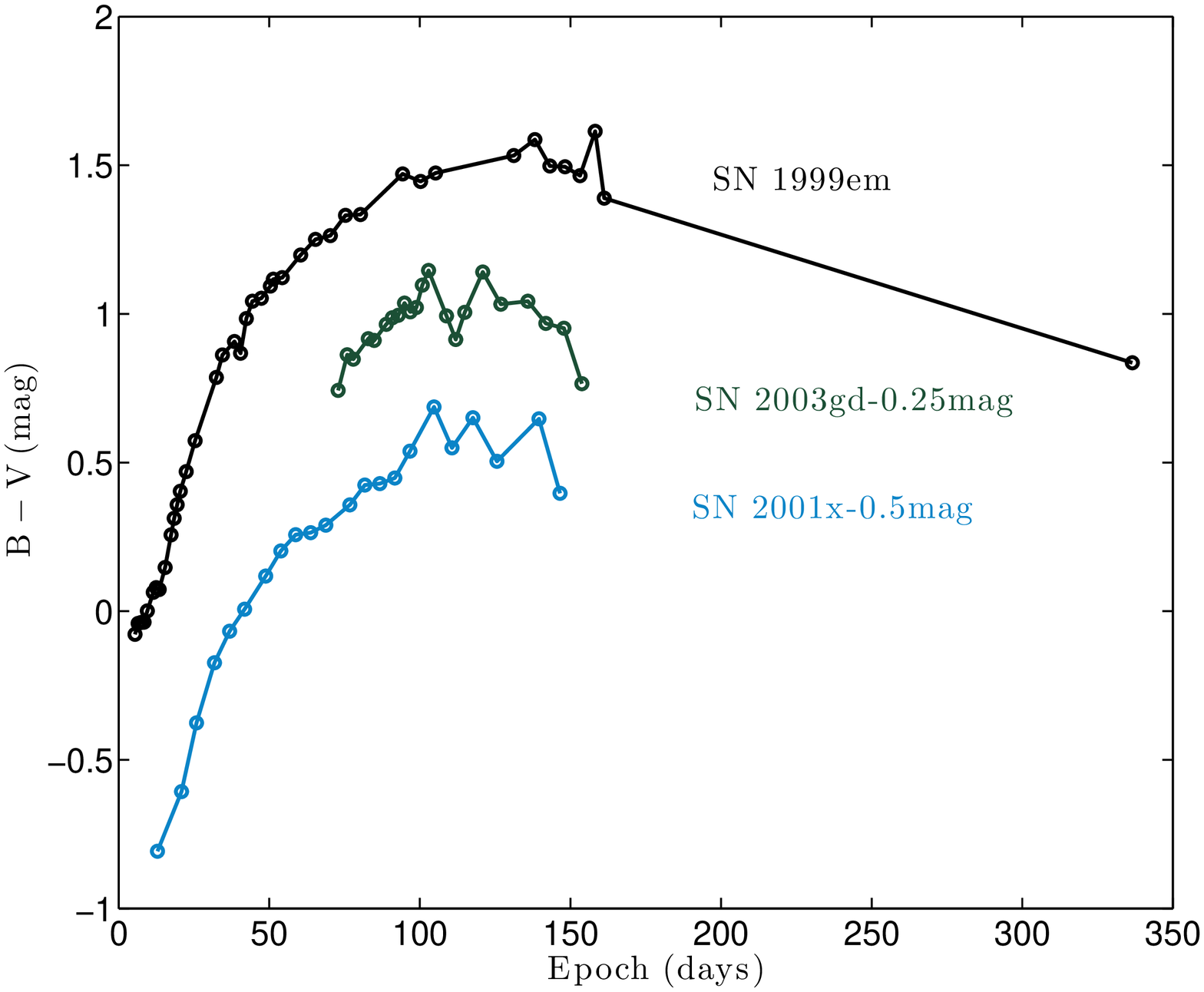}
\caption{SNe for which we have post-plateau measurements exhibit a phase of rising $B-V$ colour after the end of the plateau. This behaviour was predicted by \citet{Chieffi:2003}, and it might be the result of the rising helium abundance in the photosphere, which decreases the recombination temperature and causes a more rapid release of energy.}
\label{f:BV_rise}
\end{figure}

At the end of the plateau phase, for the objects for which we have extending post-plateau data (SNe\,1999em, 2001X, and 2003gd) one can see a rebrightening in $B-V$ as shown in Figure \ref{f:BV_rise}. In SNe\,2001X and 2003gd we also see a corresponding $B$-band brightening. This can be interpreted to result from the rising helium abundance in the photosphere, which decreases the recombination temperature and causes a rapid release of energy \citep{Chieffi:2003}.

\subsection{Rise Time}

\begin{figure}
\centering
\includegraphics[width=1\columnwidth]{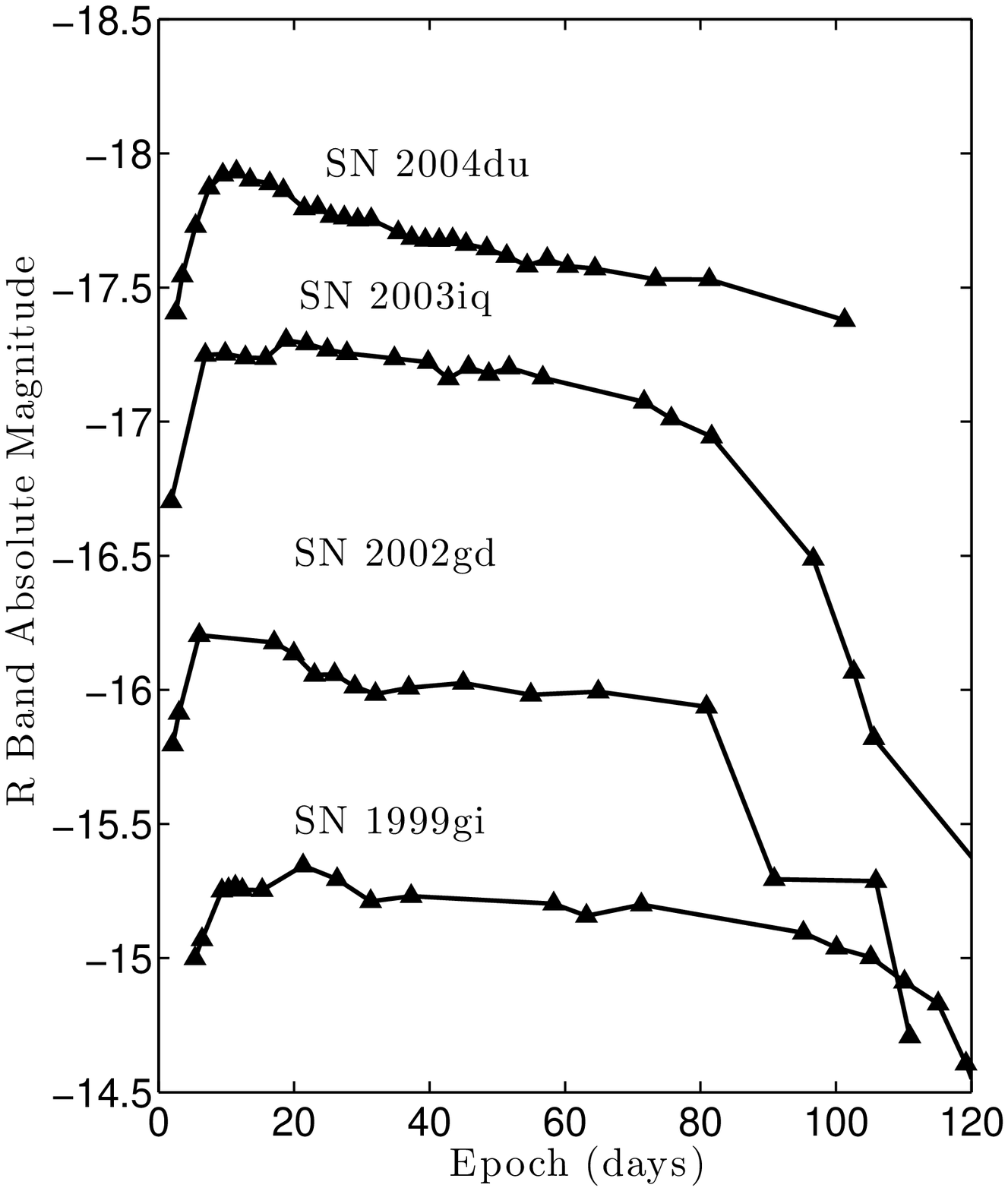}
\includegraphics[width=1\columnwidth]{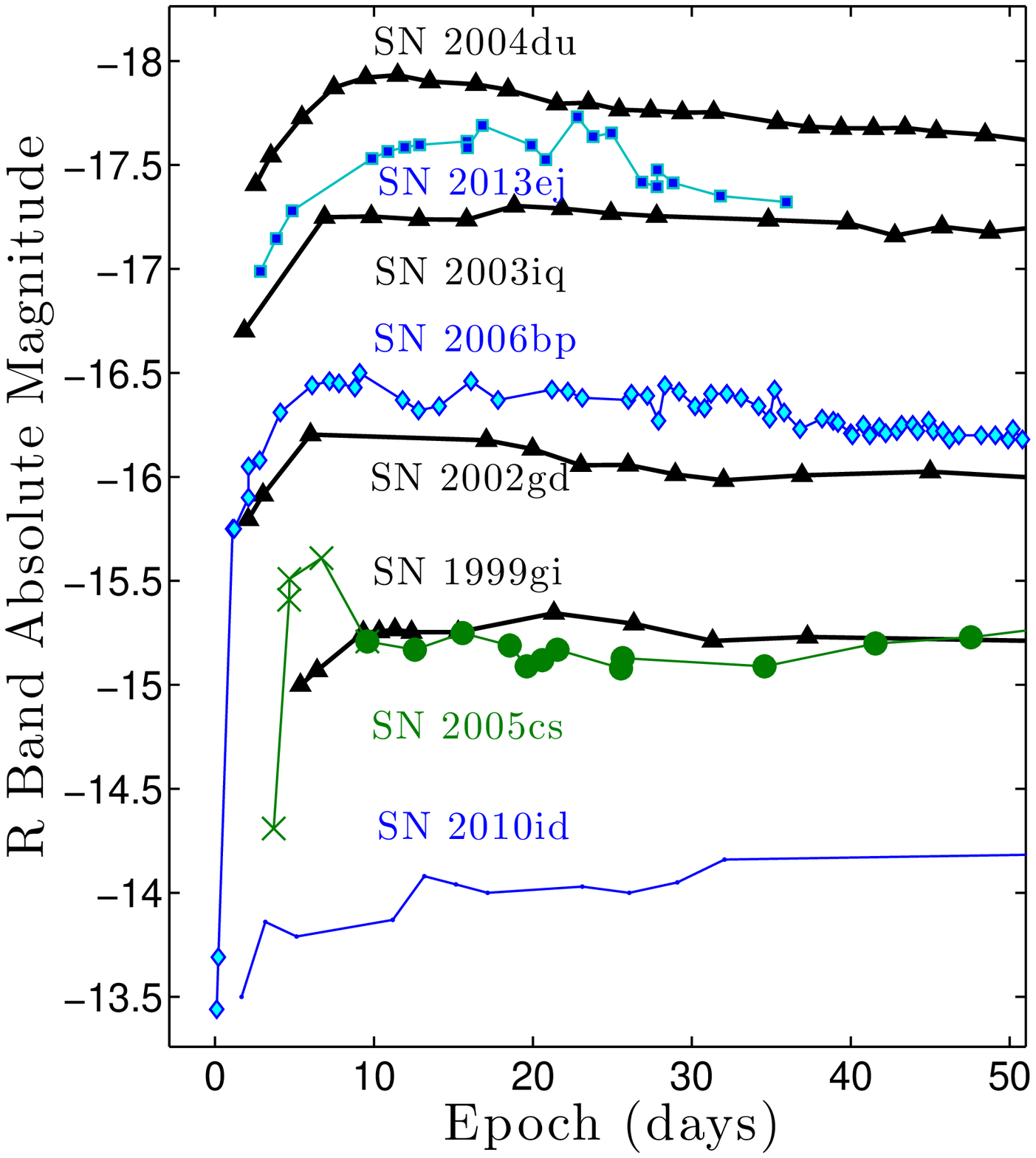}
\caption{Upper Panel: $R$-band light curves of the four events for which we have captured the early rise to the plateau. SN\,2004du is the only one that also has an early bump peaking $\sim 12$ days past explosion. Lower panel: The early $R$-band light curves of 8~SNe. The rise time and the absolute magnitude on the plateau do not seem to correlate.}
\label{f:risetime}
\end{figure}

It is usually hard to detect a SN~II-P during its early rise to maximum brightness owing to the rapid nature of this stage, so light curves usually begin on the plateau. Luckily, four of our SNe~II-P were detected sufficiently early to reveal part of the rise, and we display their $R$-band light curves in Figure \ref{f:risetime}. Our data show that most of the rise time to maximum brightness occurs within a handful of days, after which the emission settles onto the plateau within a few more days; however, there are significant variations in the light-curve shapes. 

\citet{GalYam:2011} suggest a correlation between the rise time of SNe~II-P and their luminosity. They show that two underluminous events (SN\,2005cs and SN\,2010id) rise sharply to the plateau, compared to the more luminous SN\,2006bp. To the four SNe in our sample for which we have early-time data, we add the recent SN\,2013ej from \citet{Valenti:2013} and the events used by \citet{GalYam:2011}: SN\,2006bp \citep{Quimby:2007} and SN\,2010id \citep{GalYam:2011}. We also add from \citet{Tsvetkov:2006} data for SN\,2005cs, in order to cover the earliest parts of the light curve, as done by \citet{GalYam:2011}. The early-time data for SN\,2005cs were collected by amateur astronomers and were taken from the Astrosurf website\footnote{http://www.astrosurf.com/snweb2/2005/05cs/05csMeas.htm}. Accordingly, the early peak in the light curve of this SN should be considered with caution. 

Figure \ref{f:risetime} shows the $R$-band evolution of this largest sample ever compiled of early-time SN~II-P light curves. Upon visual inspection, there does not seem to be any correlation between rise time and luminosity. A quantitative assessment will be inconclusive, owing to the still-small sample size, the sparse sampling of the light curves, and difficulty defining the endpoint consistently. 

The plateau starts once the temperatures drop below $\sim 8000$~K. It implies, based on Equation 31 of \citet{Nakar:2010}, that the time to the plateau (roughly the rise time) is $t_{\rm rise} \propto M^{-0.23}\, E^{0.2}\, R^{0.68}$. Taking $E \propto M^3$, $M \propto v$, and $L \propto v^2$ \citep{Poznanski:2013} implies $t_{\rm rise} \propto L^{0.18} R^{0.68}$. As we see, the dependence of the rise time on luminosity is weak, and most of the variance would come from differences in radii.

\section{Spectroscopic comparison}

\begin{figure*}
\centering
\includegraphics[scale=0.2]{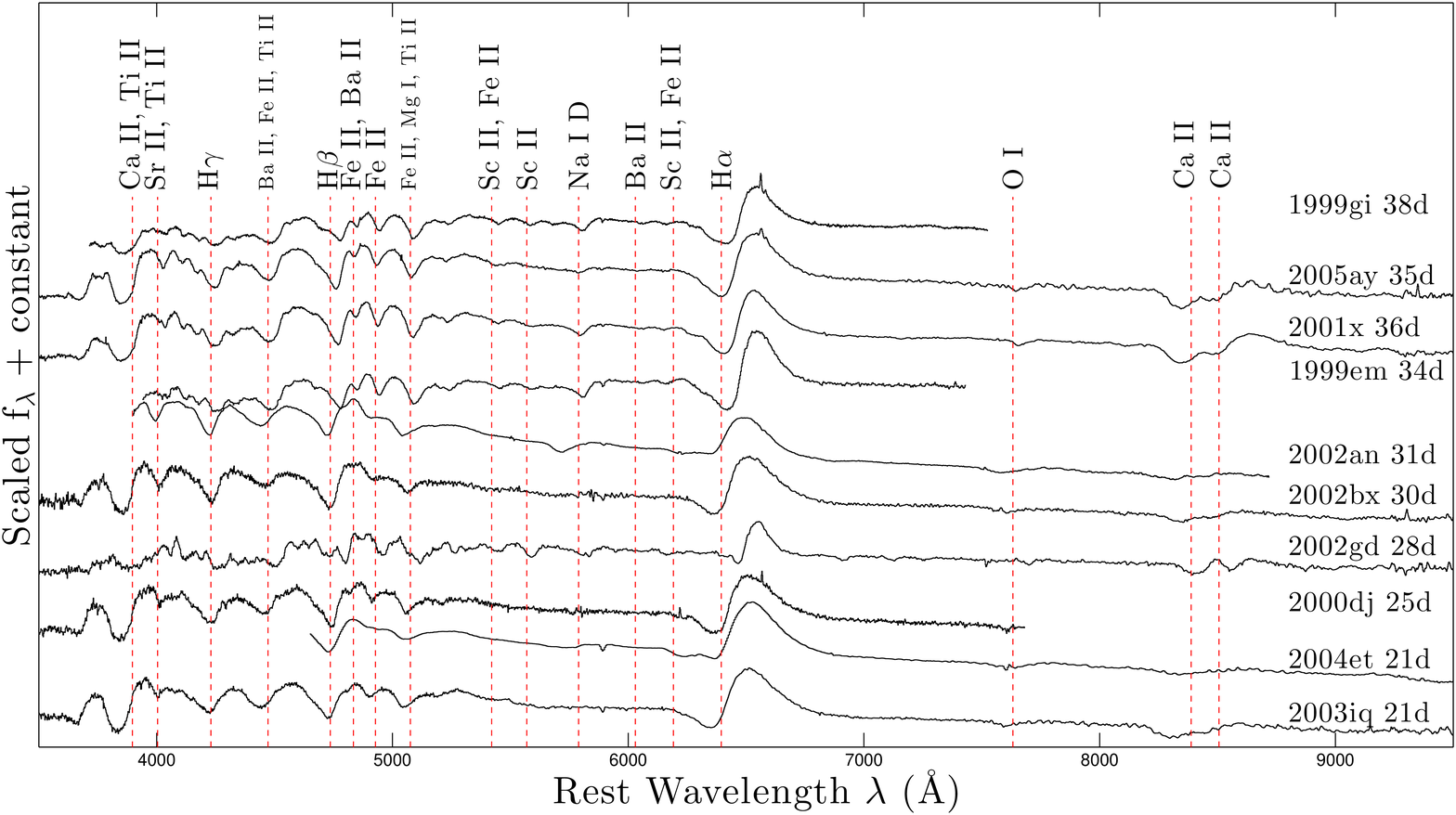}
\includegraphics[scale=0.2]{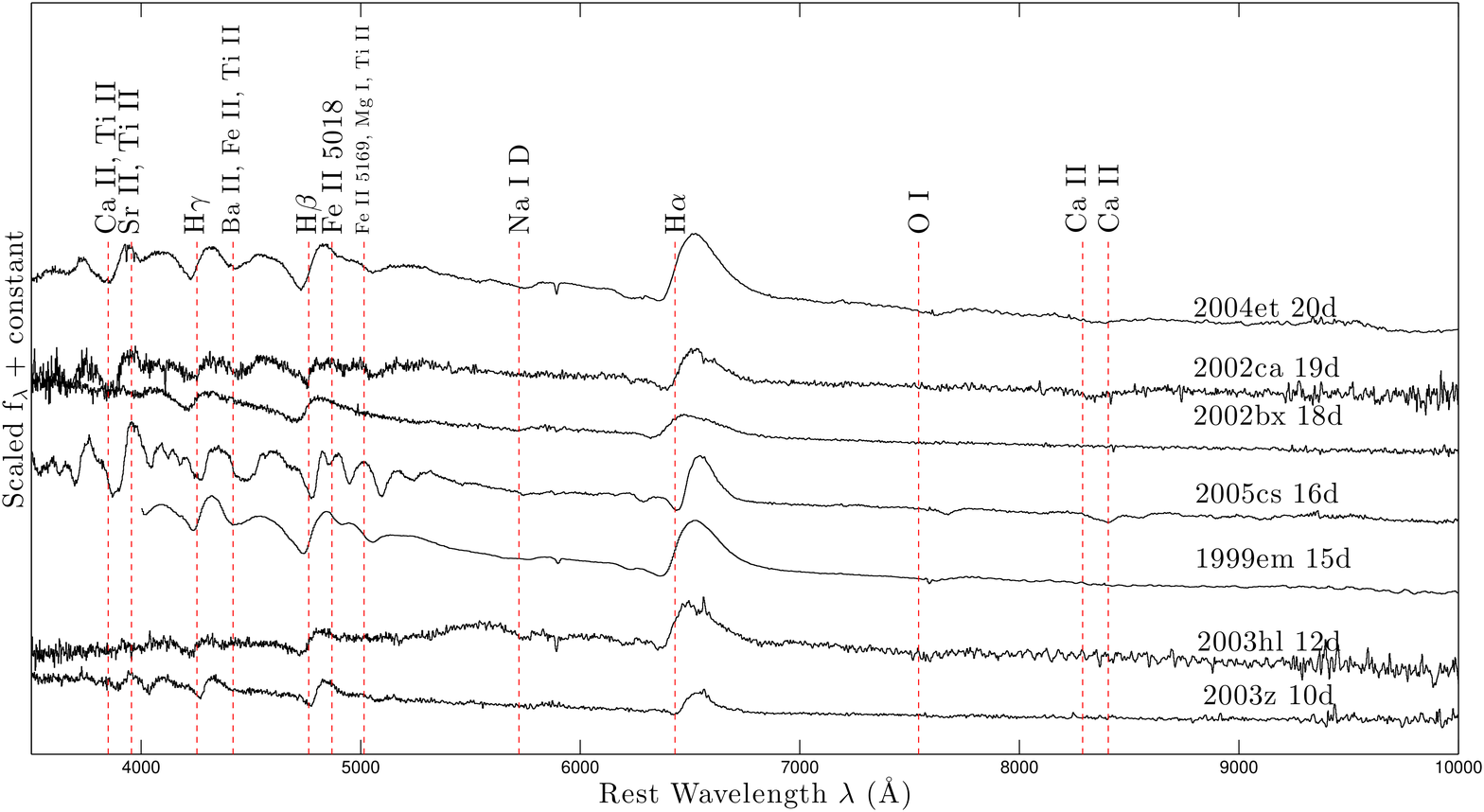}
\includegraphics[scale=0.3]{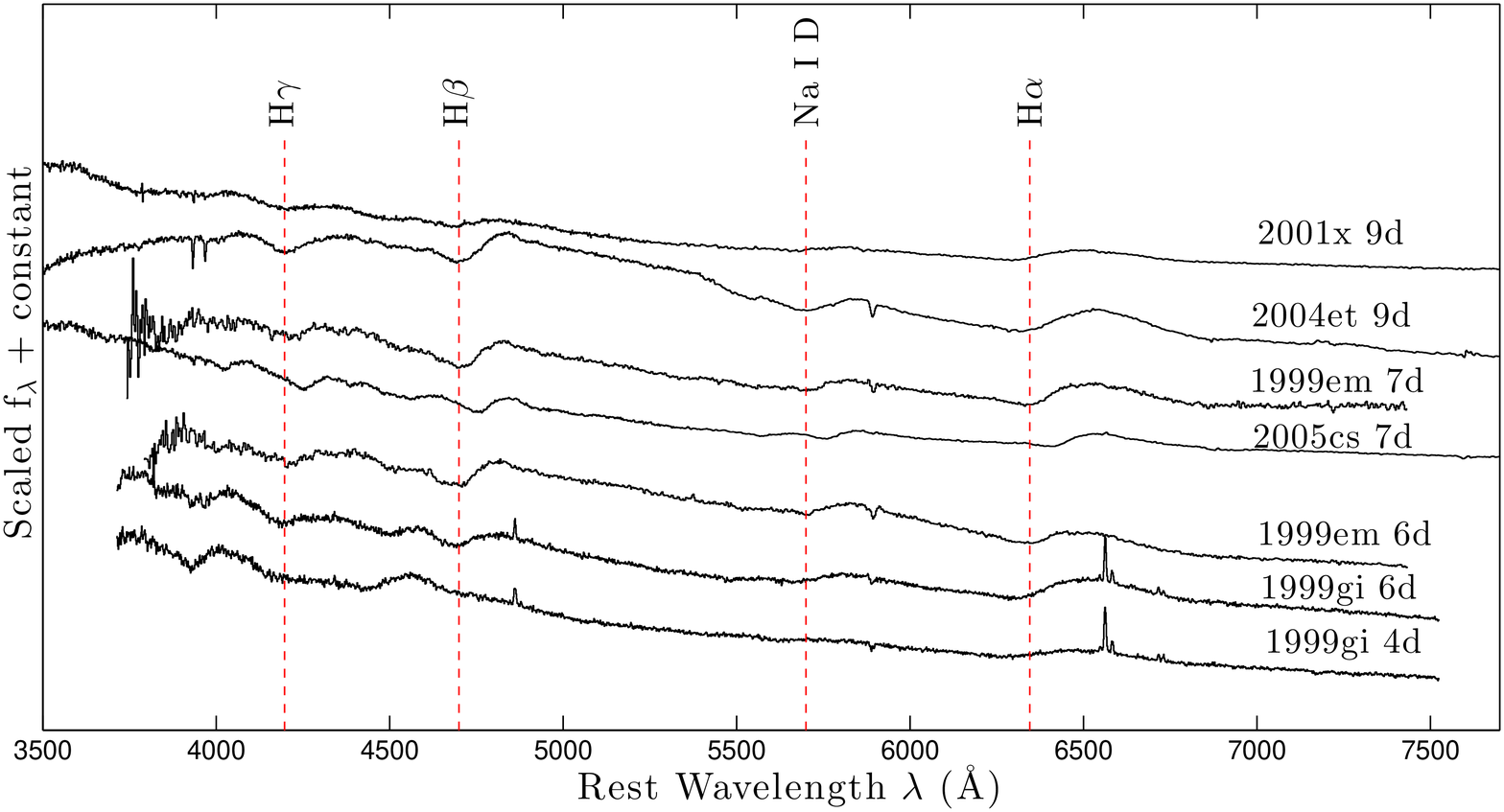}

\caption{Representative spectra 0--40 days after explosion. Dashed vertical lines indicate the wavelengths of the specified species displaced by the mean Fe~II velocities during those epochs, and by the mean H$\alpha$ velocities for Balmer lines. Early spectra before day 10 are very hot and display only Balmer (H$\alpha$, H$\beta$, and H$\gamma$) and He~I $\lambda$5876 P-Cygni profiles. Between days 10 and 20, the spectra flatten and lines of heavier species appear. All absorption and emission lines are very broad due to high velocity dispersion. Absorption lines are enhanced as well as emission lines towards mid-plateau phases. As the SN approaches the nebular phase and temperatures drop, one can see reduced absorption and an enhancement in the line emission.}
\label{f:spectra1}
\end{figure*}
\begin{figure*}
\centering
\includegraphics[scale=0.2]{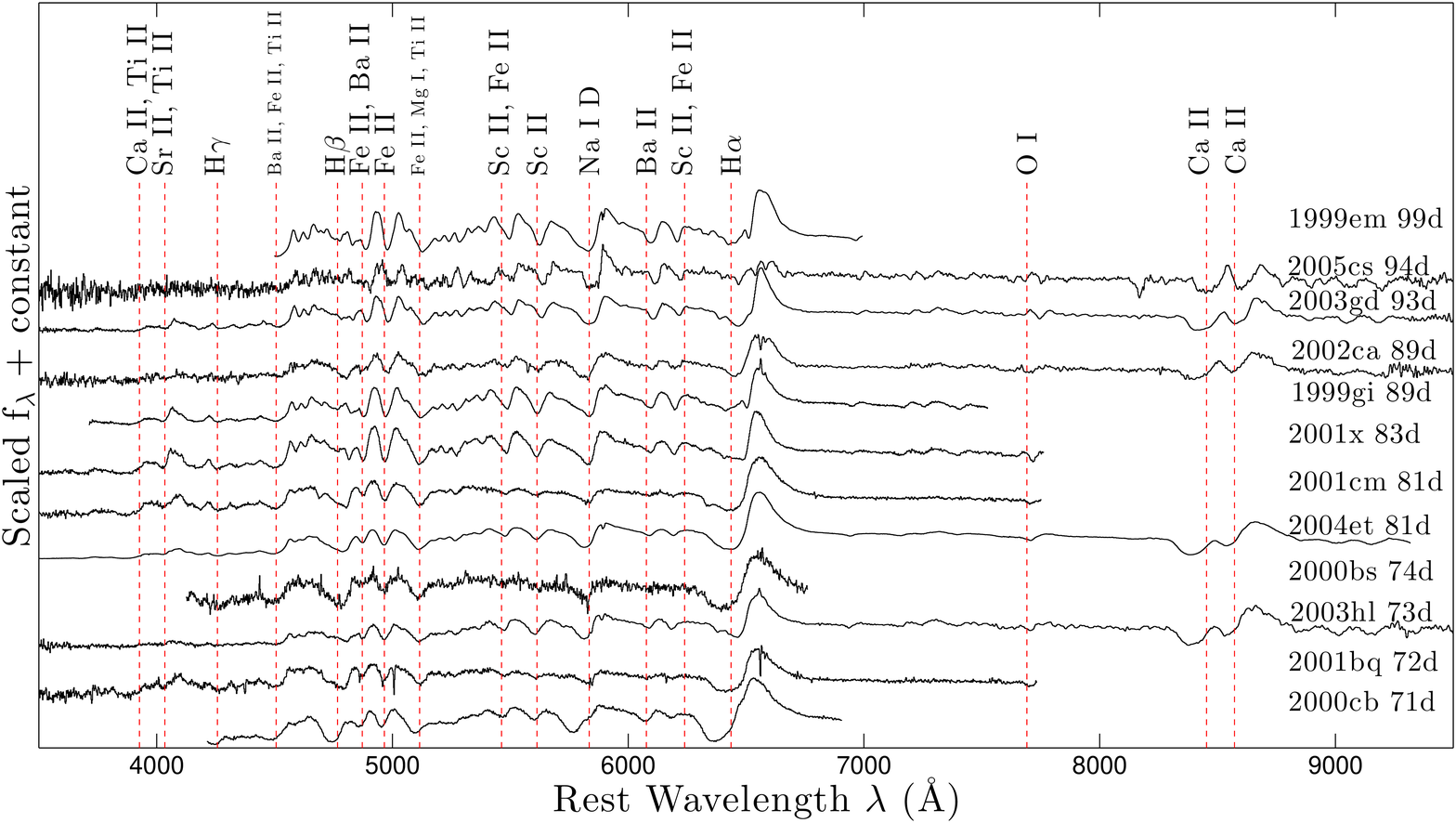}
\includegraphics[scale=0.2]{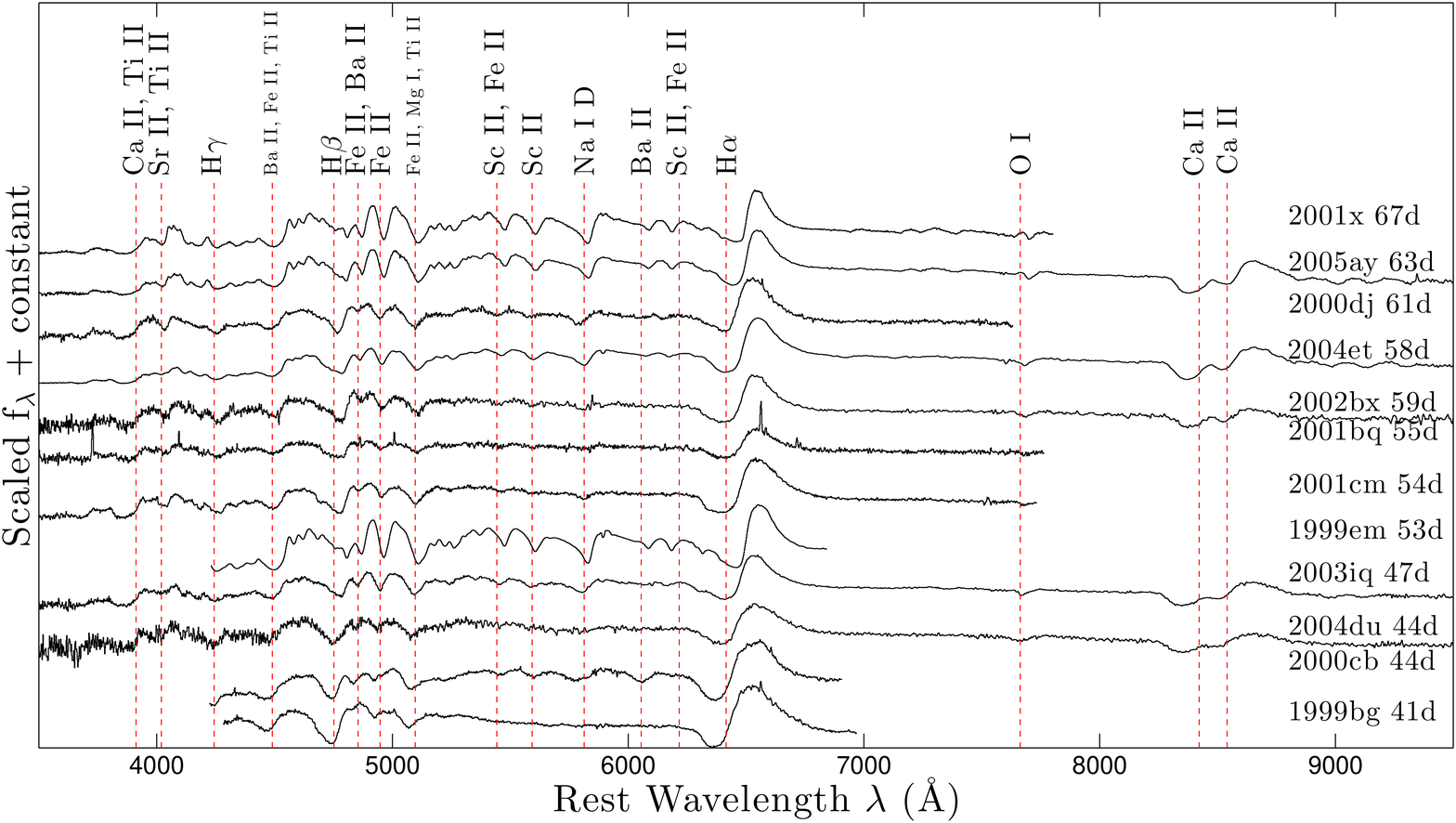}
\caption{Same as Figure \ref{f:spectra1}, for 40--100 days past explosion. Absorption and emission lines are clearly stronger towards mid-plateau phases. As a SN approaches the nebular phase and temperatures drop, one can see reduced absorption and an enhancement of line emission.}
\label{f:spectra2}
\end{figure*}

In Figures \ref{f:spectra1} and \ref{f:spectra2} we show the spectral evolution of our sample of SNe~II-P until the end of the plateau phase. In each panel we plot a representative subsample of our spectra in the indicated range of epochs including line identifications. The spectral lines were identified using the local thermodynamic equilibrium (LTE) synthetic spectra of \citet{Hatano:1999}. We compared the synthetic spectrum of various ions to our spectra, looking for coinciding lines. Candidate ions were those whose optical depth is close to unity at the temperature of the photosphere, which we assumed to be in the range of 6000--10,000~K. Once a feature was associated with a specific species, we searched for other strong lines of the same ion in the spectrum in order to confirm or reject the identification. When several species blend, we order them by the importance of their contribution. 

The first panel of Figure \ref{f:spectra1} displays the early hot days of the SNe, where the spectra are very blue and only few lines are present. Balmer (H$\alpha$, H$\beta$, and H$\gamma$) and the Na~I~D (blended with He~I) lines exhibit prominent and broad P-Cygni profiles, with blueshifted emission components in the H$\alpha$ and H$\beta$ profiles. 

\begin{figure*}
\centering
\includegraphics[width=1\textwidth]{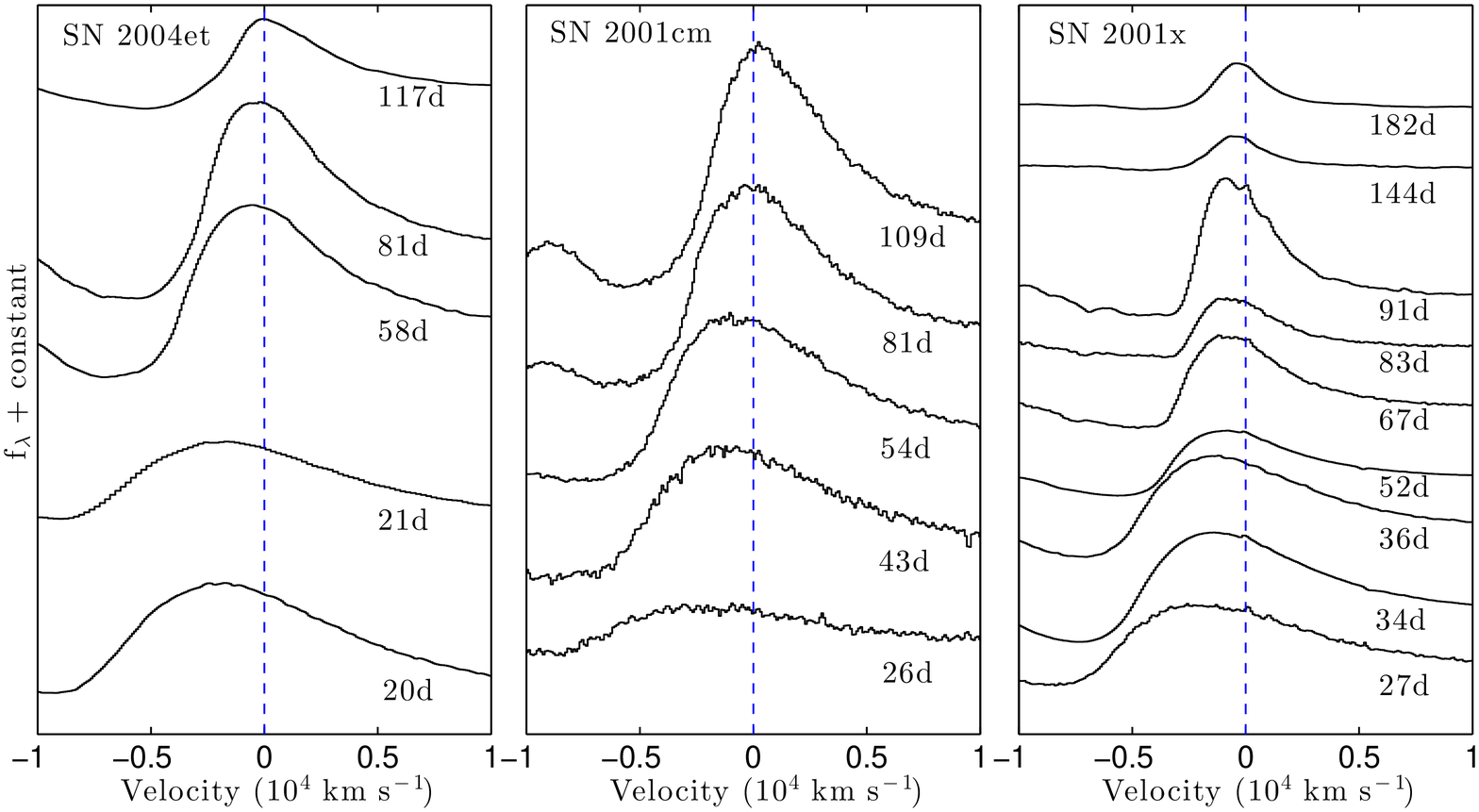}
\includegraphics[width=1\textwidth]{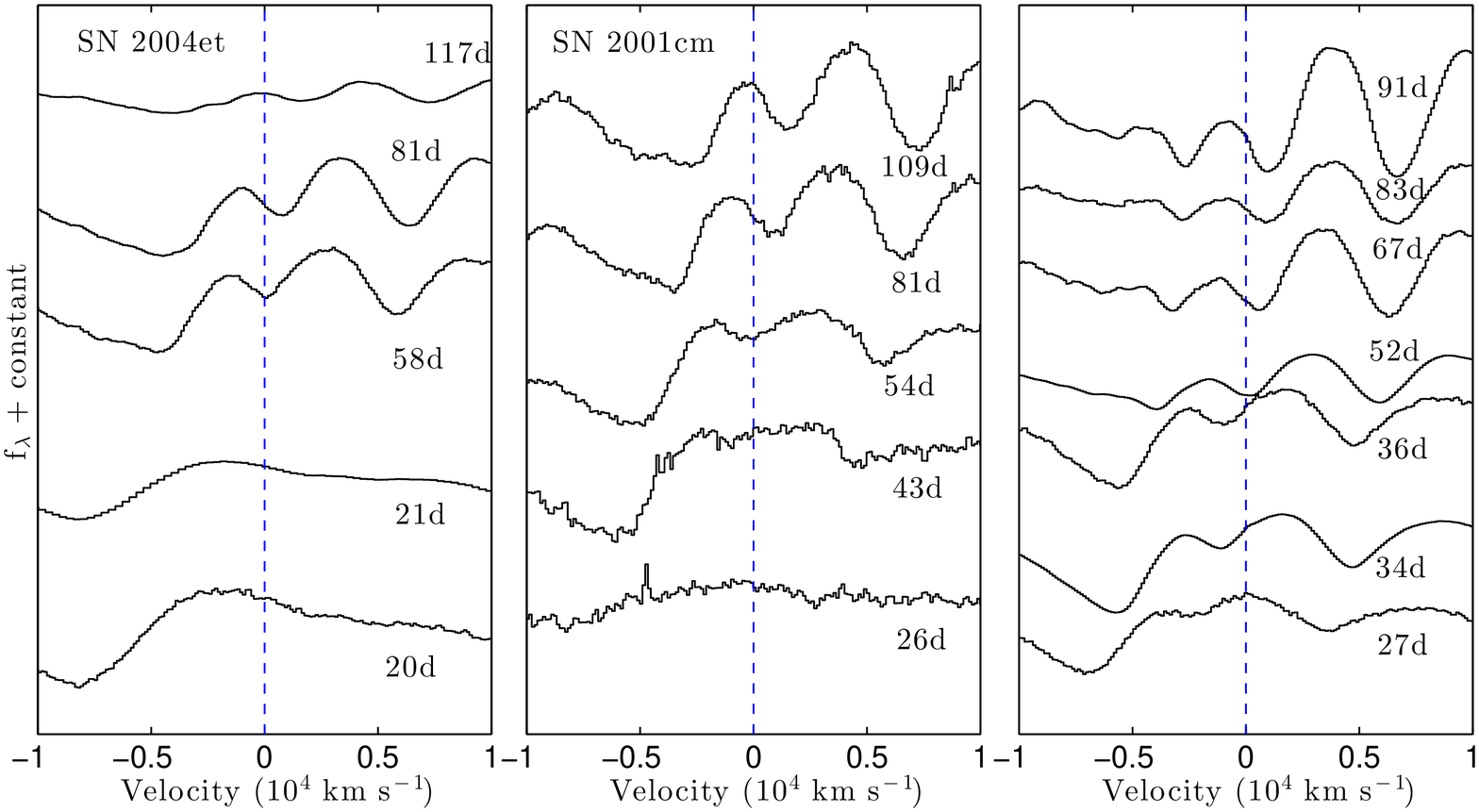}
\caption{P-Cygni profiles of H$\alpha$ (upper panel) and H$\beta$ (lower panel) for three SNe: SN\,2004et, SN\,2001cm, and SN\,2001X. The emission component is  blueshifted by $\sim 2000$~km~s$^{-1}$ in early spectra (around day 20), but as the SN evolves with time the shift decreases, and it almost disappears when the SNe approach the nebular phase.}
\label{f:blueshift}
\end{figure*}

In Figure \ref{f:blueshift} we illustrate the evolution of H$\alpha$ and H$\beta$ for three representative SNe (SN\,2004et, SN\,2001cm, and SN\,2001X). The absorption and emission features are very broad owing to large dispersion in the ejecta velocities. The absorption full width at half-maximum intensity is $\sim 6000$~km~s${-1}$ around day 25. The deviation of the emission peak from the rest wavelength is substantial during the first days and reaches $\sim$2000~km~s$^{-1}$ around day 20, after which it declines. On day 57 the shift is only $\sim 540$~km~s$^{-1}$ for SN\,2004et and it completely disappears by day 117. SN\,2001cm behaves similarly, whereas SN\,2001X is still blueshifted by $\sim 250$~km~s$^{-1}$ on day 181. The assumption of a spherically symmetric expansion should lead to the formation of a symmetric emission profile around zero velocity, in contrast to what we actually see. \citet{Dessart:2005} explain these observations using the CMFGEN model atmosphere code \citep{Hillier:1998}, which solves the radiative transfer and statistical equilibrium equations in an expanding medium, assuming radiative equilibrium. They show that the blueshift of the emission peak arises from the combined effect of disk occultation and a line-forming region very close to the photosphere. The confinement of that region to the photosphere causes a significant flux deficit in the red side of the emission profile, which is enhanced for rapidly declining density profiles, like the ones assumed for CC-SNe. The offset of the emission peak from zero velocity declines with time, as the occulted region becomes smaller with the photosphere's recession into the star. 

By day 10 the spectra flatten, and around day 15, Fe~II $\lambda$5169 and the Ca~II lines emerge along with O~I $\lambda$7774. By day 20 we can already see a strong line of  Ba~II $\lambda$4554 (probably blended with Fe~II and Ti~II) and a hint of what appears to be Sr~II $\lambda$4077 (blended with Ti~II). 

During this range of epochs, most spectra are not yet very developed. However, SN\,2005cs shows many details quite early relative to other objects (e.g., see the spectrum at day 16 in Figure \ref{f:spectra1}). Around day 10 a notch appears in the H$\alpha$ blue wing of several SNe. This feature, possibly due to hydrogen at high velocity (HV), gets stronger and remains visible until the end of the plateau phase. We will discuss this matter in detail in Section \ref{s:HV}. 

A few more lines appear near day 35, where we can perhaps see Ba~II $\lambda$6141 and Sc~II $\lambda\lambda$6305, 5672, 5520, which are strongest for SN\,2005cs. On the second half of the plateau and toward its end, the strong  Balmer H$\beta$ and H$\gamma$ lines, as well as Ca~II H\&K, become wider and less distinct owing to blending with (perhaps) Ti~II and Fe~II, and are thus much more difficult to detect. One can also notice the growing emission lines along with the reduction of the absorption components, as the SNe approach the nebular phase.

The absorption EW of the hydrogen lines grows approaching mid-plateau epochs, and declines towards the nebular phase. The general shape of the evolution can be interpreted as follows. When the SN is young and hot, the hydrogen is fully ionised and its optical depth at the photosphere is close to 1, generating only little absorption. When the photosphere cools and hydrogen atoms start recombining with electrons, the optical depth approaches $\sim 1000$, and simultaneously more material is revealed as the photosphere recedes, bringing the line profile to maximal absorption. By the end of the plateau phase, the density is small enough to increase the probability of photon escape without going through absorption in hydrogen atoms, and the EW decreases slowly.

\subsection{Ejecta Velocities}
\label{s: Photospheric and Ejecta Velocities}

We measure the velocities of H$\alpha$, H$\beta$, and Fe $\lambda$5169 in every spectrum where they can be detected. We do so by fitting a third-order polynomial to the absorption component of the P-Cygni line, and taking its minimum as the center of the velocity distribution. We do not rectify the lines owing to the difficulty in defining the continuum around the profiles, which might cause systematic errors in our measurements. In order to estimate measurement uncertainties, we smooth the data using a Savitzky-Golay filter \citep{Savitzky:1964}. We then divide our spectra by their smoothed versions, thus generating a vector of noise. We randomly mix the noise vector in wavelength to create $10^{4}$ different vectors. Each artificial noise spectrum is then multiplied by the smoothed spectrum to create a ``new'' spectrum of signal plus noise, and the velocity is measured for every one of these. The standard deviation of the results gives us the uncertainty in the original velocity measurement, and is typically $\sim 70$~km~s$^{-1}$. An uncertainty of 250~km~s$^{-1}$ is added in quadrature to account for peculiar velocities. We linearly interpolate (or extrapolate) the velocities of each object to day 50 after explosion and normalise the measurements by that value.

H$\alpha$ and H$\beta$ are detected at high velocities in early epochs and are emitted from the outer ejecta. Lower velocity layers are revealed as the photosphere recedes and lines of heavier ions emerge. The velocity of the photosphere can be traced using the Fe~II triplet, which is believed to be generated at the thermalisation surface, and is therefore considered a good proxy for the velocity near the photosphere (\citealt{Schmutz:1990}; \citealt{Dessart:2005b}). 

\begin{figure}
\centering
\includegraphics[width=1\columnwidth]{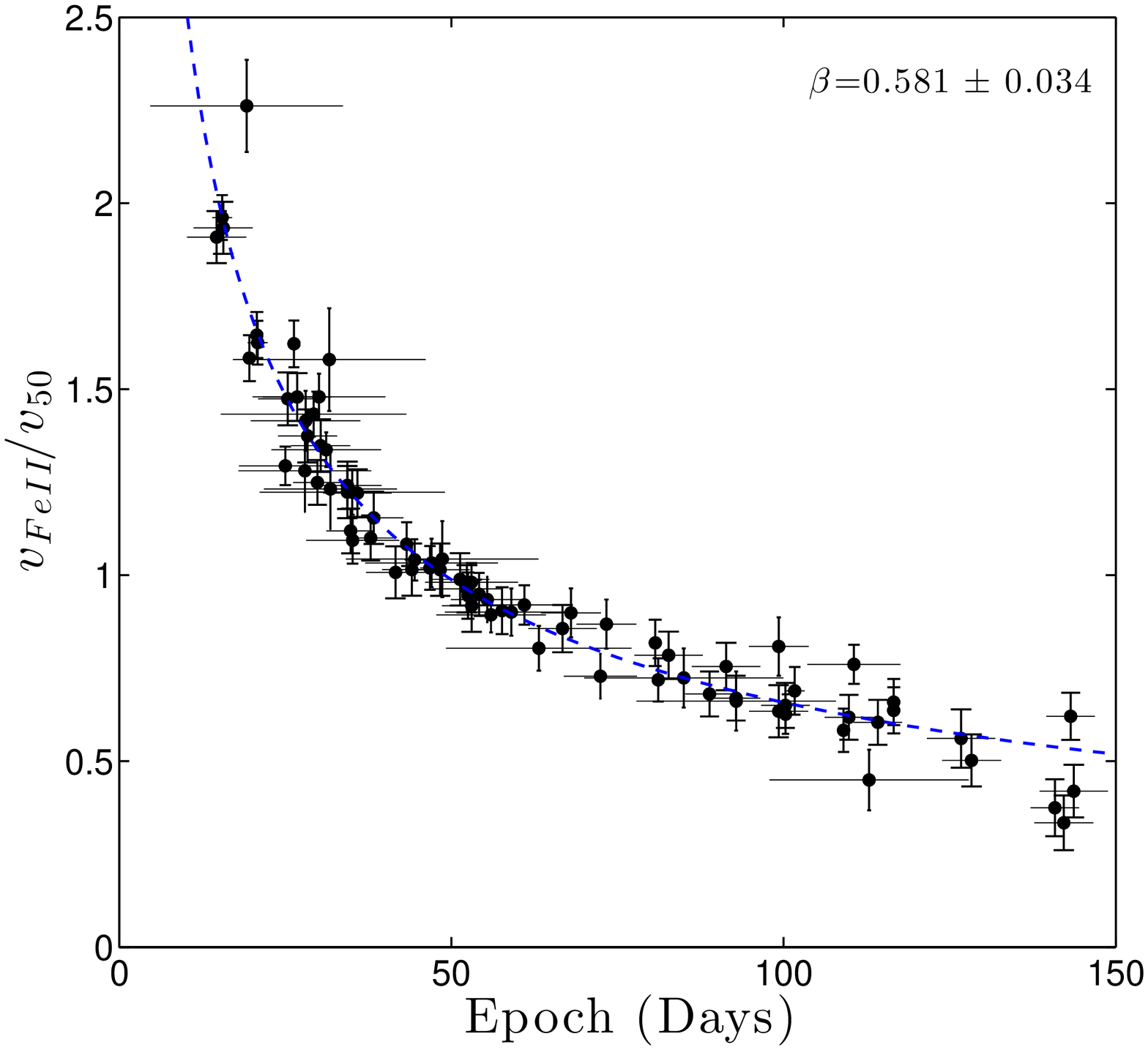}
\includegraphics[width=1\columnwidth]{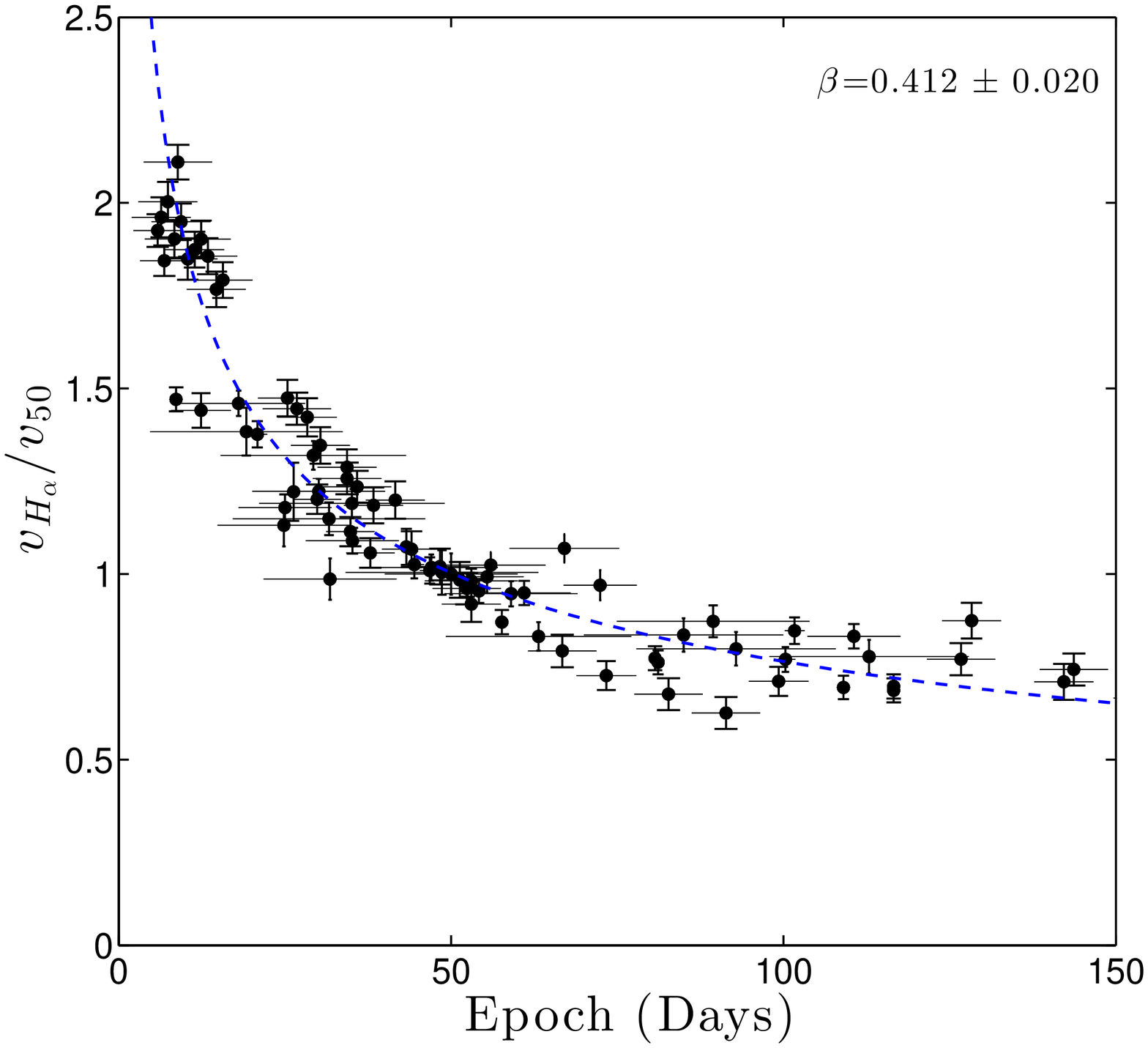}
\includegraphics[width=1\columnwidth]{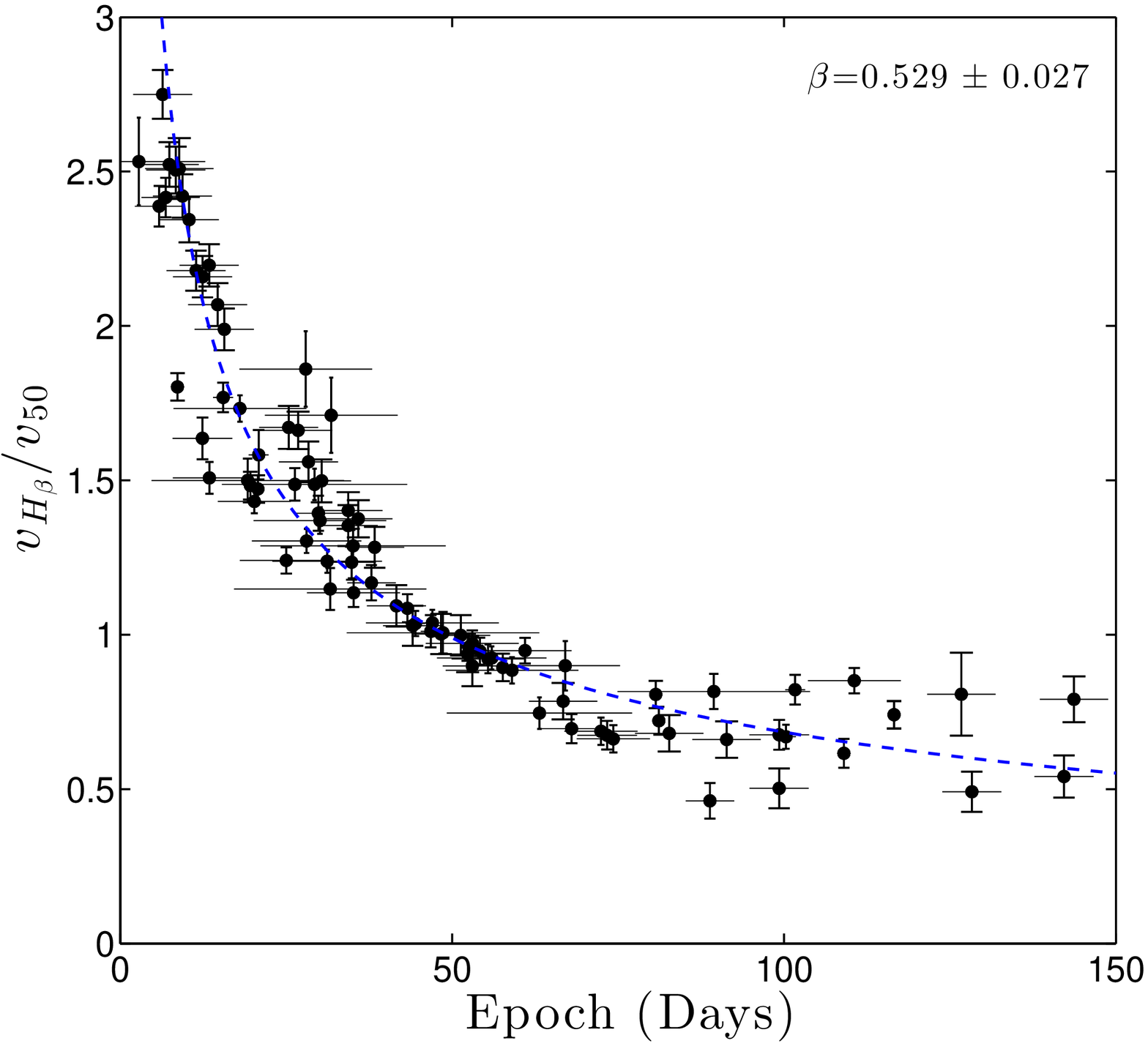}
\caption{Fe~II $\lambda$5169 (top panel), H$\alpha$ (middle), and H$\beta$ (bottom) velocities normalised by $v_{\rm ph,50}$. Best-fitting power laws are plotted with dashed lines. All SN~II-P velocities follow their respective power laws quite well and with little scatter.}
\label{f:v_t}
\end{figure}

The temporal evolution of the various scaled velocities is well fit by power laws as can be seen in Figure \ref{f:v_t}, where we use only spectra obtained before day 150. Uncertainties in the explosion date are also taken into account in the calculation. The Fe~II velocity declines with a power of $\beta= -0.581\pm 0.034$, a somewhat sharper decline than the $-0.464 \pm 0.017$ found by \citet{Nugent:2006}. Since they used a narrower time window, we repeat our fit using only spectra obtained between days 9 and 75, but we obtain roughly the same sharp decline as we did using all of our data. The difference might be caused by our better sampling of the early days, where the decline is significant and can change the fit most effectively. Overall, we can represent the SN~II-P evolution of the velocities as $v_{\rm Fe~II}=v_{\rm ph,50}\,(t/50)^{-0.581\pm0.034}$, $v_{\rm H\alpha}=v_{\rm ph,50}\,(t/50)^{-0.412\pm0.020}$ and $v_{\rm H\beta}=v_{\rm ph,50}\,(t/50)^{-0.529\pm0.027}$.

\begin{figure}
\centering
\includegraphics[width=1\columnwidth]{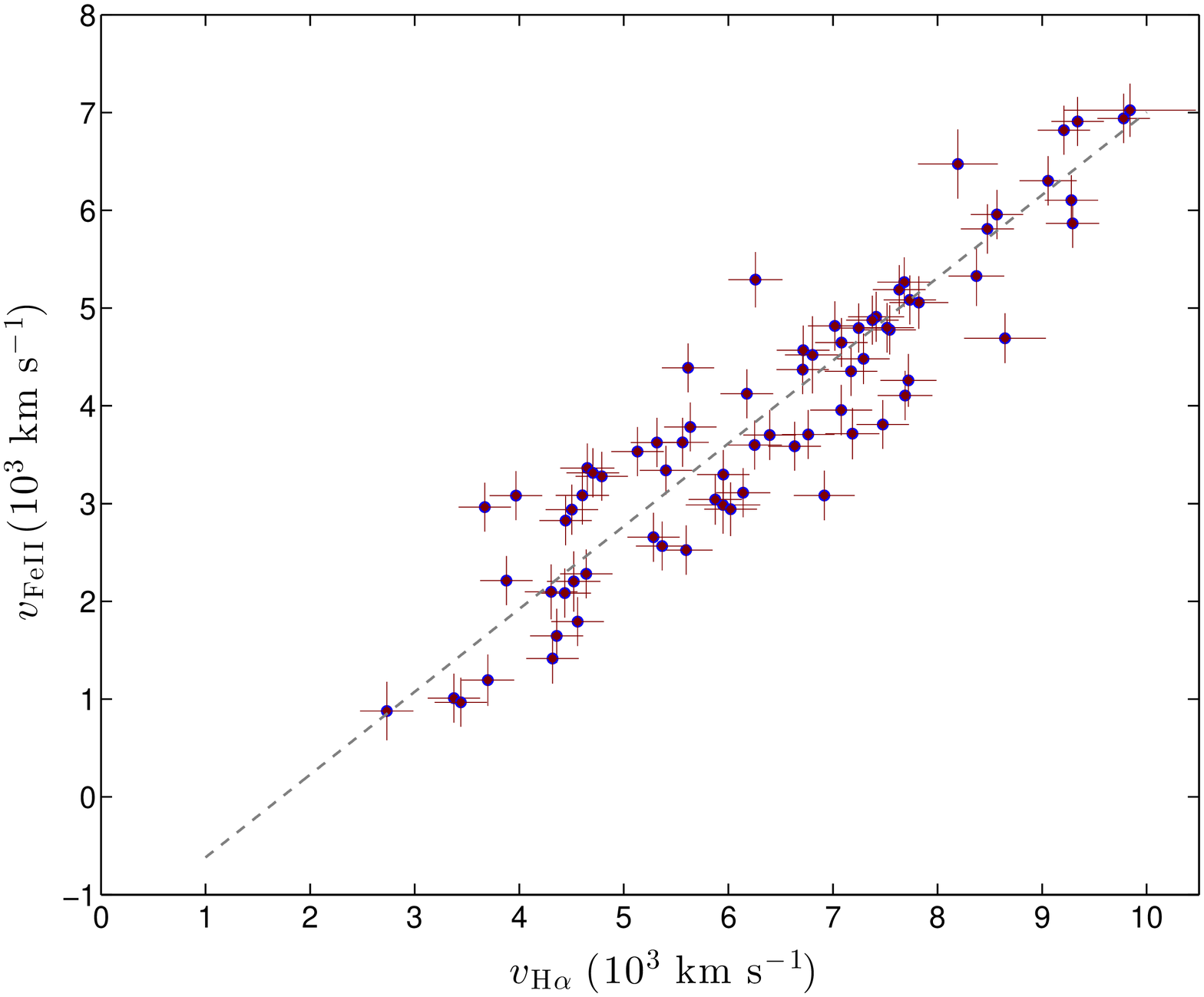}
\includegraphics[width=1\columnwidth]{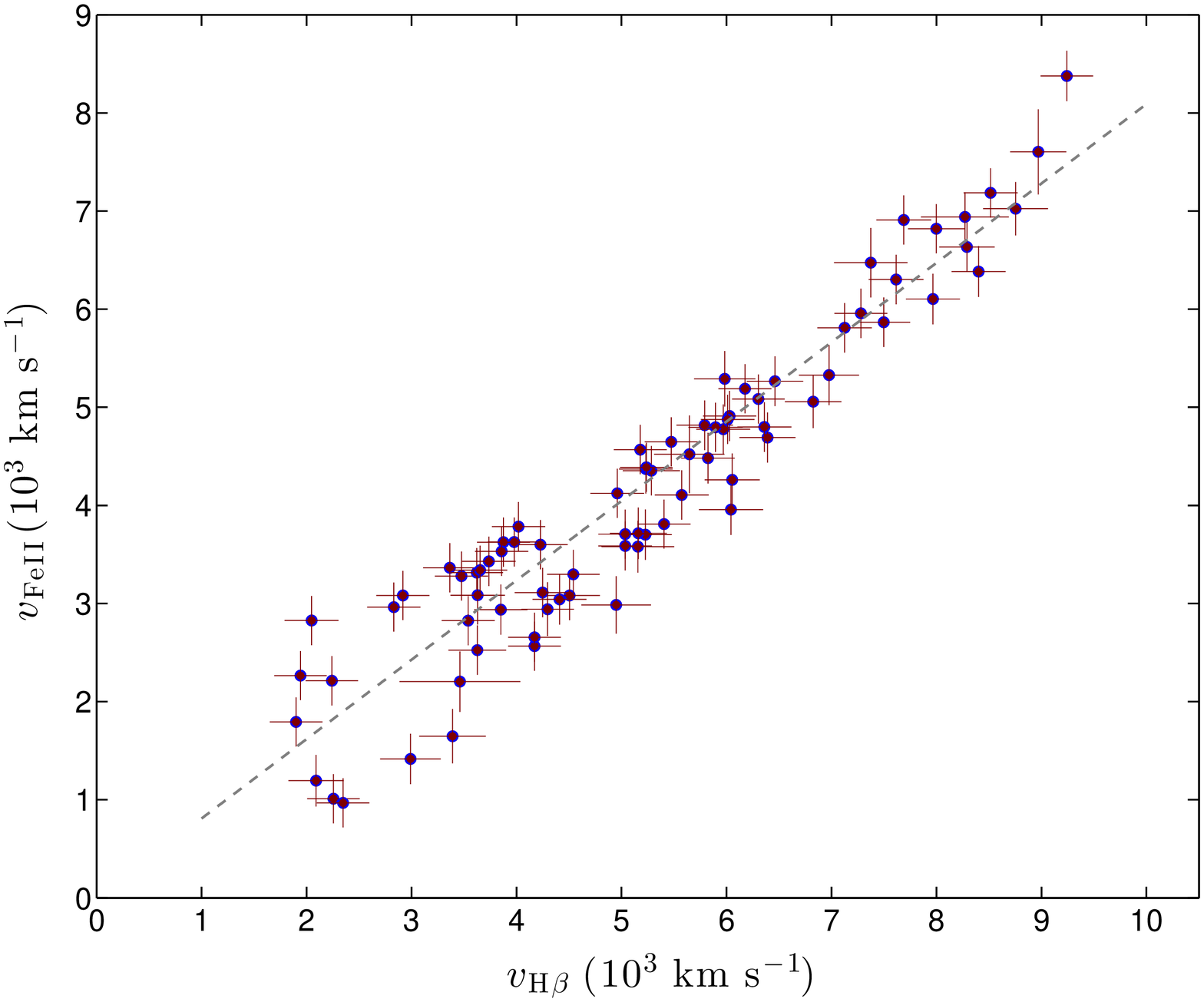}
\caption{Fe~II $\lambda$5169 velocities vs. H$\alpha$ (upper panel) and H$\beta$ (lower panel) velocities. $v_{\rm Fe~II}$  is linear with $v_{\rm H\beta}$, and the best fit gives $v_{\rm Fe~II} = (0.805 \pm 0.005)v_{\rm H\beta}$. While the $v_{\rm H\alpha}$ evolution is different from $v_{\rm Fe~II}$, their dependence can still be approximated by a linear relation, $v_{\rm Fe~II} = (0.855 \pm 0.006)v_{\rm H\alpha} - (1499 \pm 87)$~km~s$^{-1}$. }
\label{f:v_v}
\end{figure}

Our results show that $v_{\rm Fe~II}$ and $v_{\rm H\beta}$ follow a similar power law, and are thus linearly related, as was already shown by \citet{Poznanski:2010}, and earlier explored by \citet{Nugent:2006}. This allows $v_{\rm H\beta}$ to be a proxy for the photospheric velocity at early epochs, when the Fe~II lines have not yet emerged. We fit a power law of the form $v_{\rm Fe~II} = a\,v_{\rm H}^{-b}$ using 79 spectra of 20 SNe for which all the velocities are measurable. We restrict the velocities to epochs earlier than 200 days. We find that $v_{\rm Fe~II}$ is indeed linearly proportional to H$\beta$ with a power law of $b = 1.01 \pm 0.01$. Fitting a linear function we find that $v_{\rm Fe~II} = (0.805 \pm 0.006)v_{\rm H\beta}$ with ${\chi}^{2}$/dof = 1.72. This agrees with the results of \citet{Poznanski:2010}. On the other hand, $v_{\rm H\alpha}$  follows Fe~II with a power law of $b = 1.32 \pm 0.02$. Nevertheless, fitting a linear function, we find that $v_{\rm Fe~II} = (0.855 \pm 0.006)v_{\rm H\alpha} - (1499 \pm 87)$~km~s$^{-1}$ with ${\chi}^{2}$/dof = 2.23. Both fits to the data are shown in Figure \ref{f:v_v}.

\section{High-Velocity Features}
\label{s:HV}

\begin{figure*}
\centering
\includegraphics[width=0.9\textwidth]{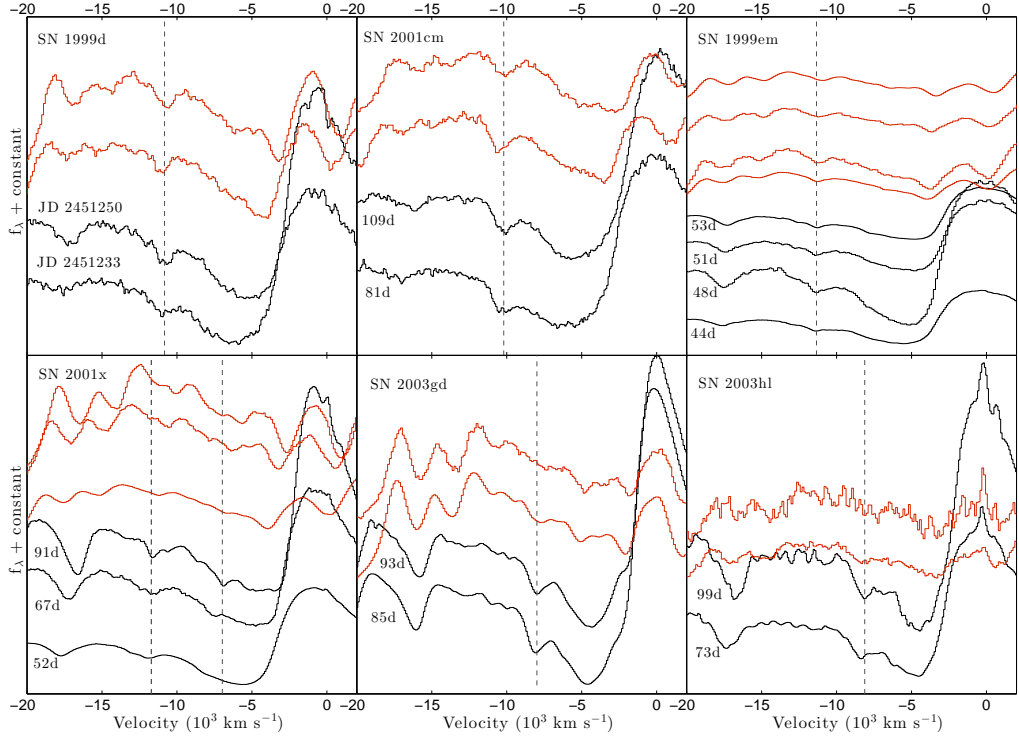}
\caption{Spectra of six SNe showing the HV features in H$\alpha$ (black) which have a counterpart in H$\beta$ (red). The HV components are seen at high velocities of $\sim -$10,000~km~s$^{-1}$ and slow down with time. The presence of matching line in H$\beta$ supports an interpretation of a hydrogen HV feature.}
\label{f:HV1} 
\end{figure*}

Eight different objects at various epochs during their photospheric phase (day 14 or earlier in SN\,1999em, day 81 in SN\,2001cm) have a blue notch in H$\alpha$, which could be high-velocity (HV) hydrogen features. Alternatively, it could be Si~II $\lambda$6355. If these are indeed HV features one should expect to find similar absorption in the profiles of other hydrogen lines. In Figure \ref{f:HV1} we compare the spectra that show the feature near H$\alpha$ (black) as well as near H$\beta$ (red). The presence of an additional absorption supports the identity of the lines as HV features, as discussed by \citet{Inserra:2012} who identified similar features in the spectra of SN\,2009bw. Note that for SN\,2001X, the velocities of the H$\alpha$ and H$\beta$ components do not match, and the identification as HV features is therefore dubious. The HV lines follow the redward drift with time like the rest of the spectral lines.

\begin{figure}
\centering
\includegraphics[width=0.95\columnwidth]{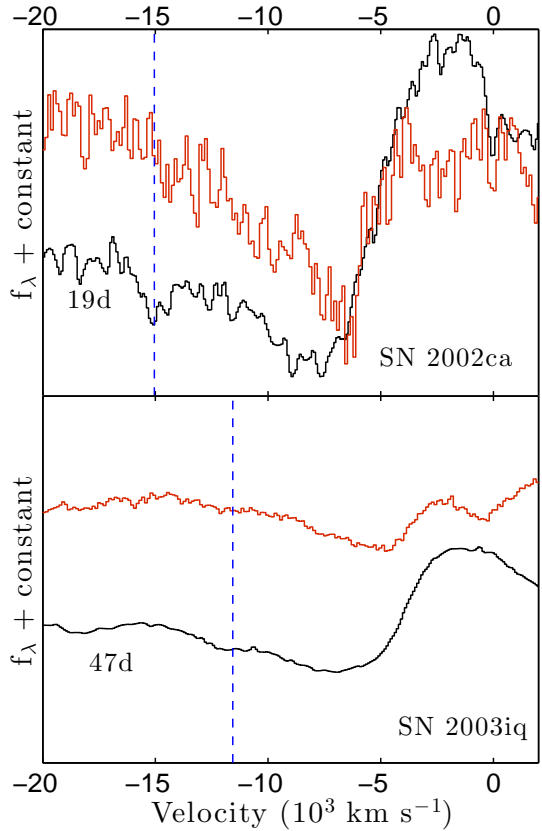}
\caption{Same as Figure \ref{f:HV1} for SNe with a blue notch in H$\alpha$ but not in H$\beta$, favouring the interpretation as Si~II $\lambda$6355.}
\label{f:HV2} 
\end{figure}

In Figure \ref{f:HV2} we show the spectra of two SNe where no counterpart is detected in H$\beta$. The spectrum of SN\,2002ca has low signal-to-noise ratio, which might affect our ability to detect the H$\beta$ component, but that is not the case for the other SN. We therefore favour an interpretation of Si~II $\lambda$6355 at a velocity of $\sim 5000$~km~s$^{-1}$, typical to metals in SNe~II. In a similar case, \citet{Valenti:2013} see a wide absorption feature in SN\,2013ej on the blue side of H$\alpha$ without a counterpart in H$\beta$, and interpret it as a line of Si~II $\lambda$6355. However, it is much stronger than the lines in our sample.

\begin{figure}
\centering
\includegraphics[width=.8\columnwidth]{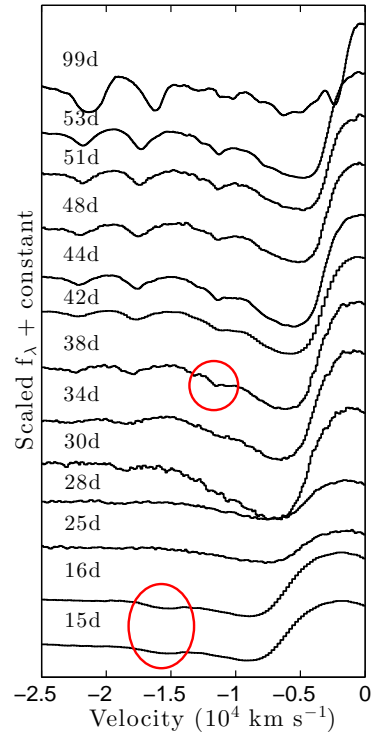}
\caption{The evolution of the HV feature in the H$\alpha$ blue wing of SN\,1999em. A HV component is evident in the first two lower spectra at a velocity of $\sim -$15,000~km~s$^{-1}$, and it disappears on day 25. It suddenly reappears in the spectrum of day 38 and remains visible until the end of the plateau phase.}
\label{sHV1999em} 
\end{figure}

It is interesting to follow the time evolution of the H$\alpha$ HV feature in the spectra of SN\,1999em. As seen in Figure \ref{sHV1999em} the feature is first evident in the ninth spectrum on day 15, and appears relatively strong and wide at a velocity of $\sim -$15,000~km~s$^{-1}$. It is still present on the subsequent day, but completely disappears 10 days later. A new HV feature appears again between days 34 and 38, but this time it is much narrower and does not resemble the previous feature in shape. The notch remains apparent at least until the SN enters the nebular phase. We do not see this behaviour in the HV feature we associated with H$\beta$, as it emerges only at days 34--38, together with the H$\alpha$ HV notch. The complete disappearance and the different shape of the line after it reappears suggests that the feature comes from two different emission regions or mechanisms dominating at different epochs. However, since no counterpart in H$\beta$ is found before days 34--38, the early feature might be associated with Si~II $\lambda$6355 and not HV hydrogen. 

\begin{figure}
\centering
\includegraphics[width=0.95\columnwidth]{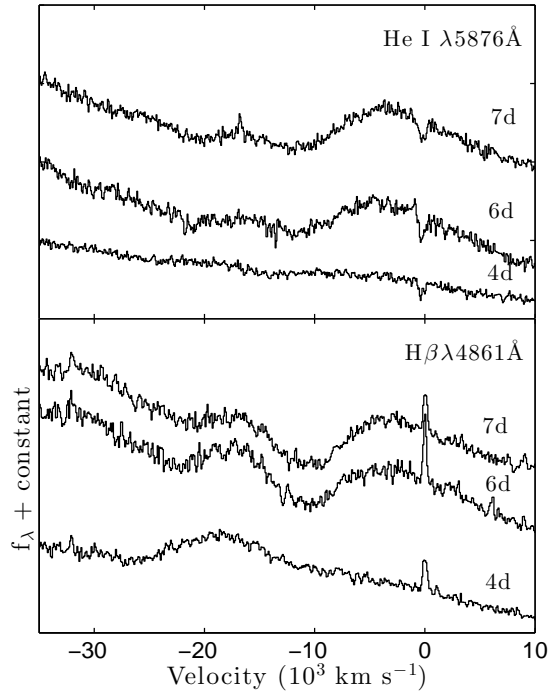}
\caption{Early-time HV features associated with H$\beta$ and He~I $\lambda$5876 in SN\,1999gi. The HV components are observed at high velocities of $\sim -$20,000~km~s$^{-1}$, with a P-Cygni profile.}
\label{sHV1999gi_early} 
\end{figure}

\begin{figure}
\centering
\includegraphics[width=0.95\columnwidth]{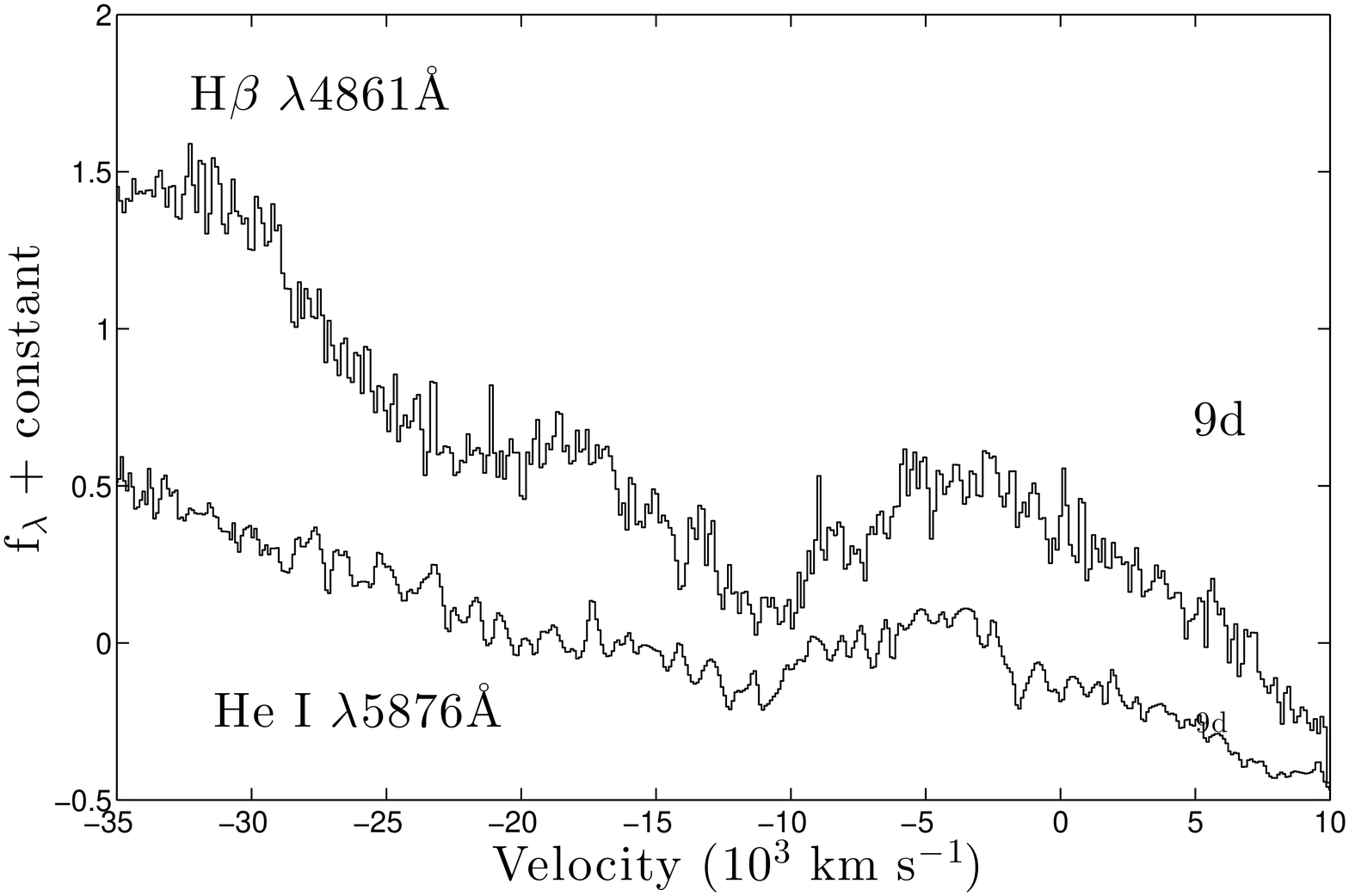}
\includegraphics[width=0.95\columnwidth]{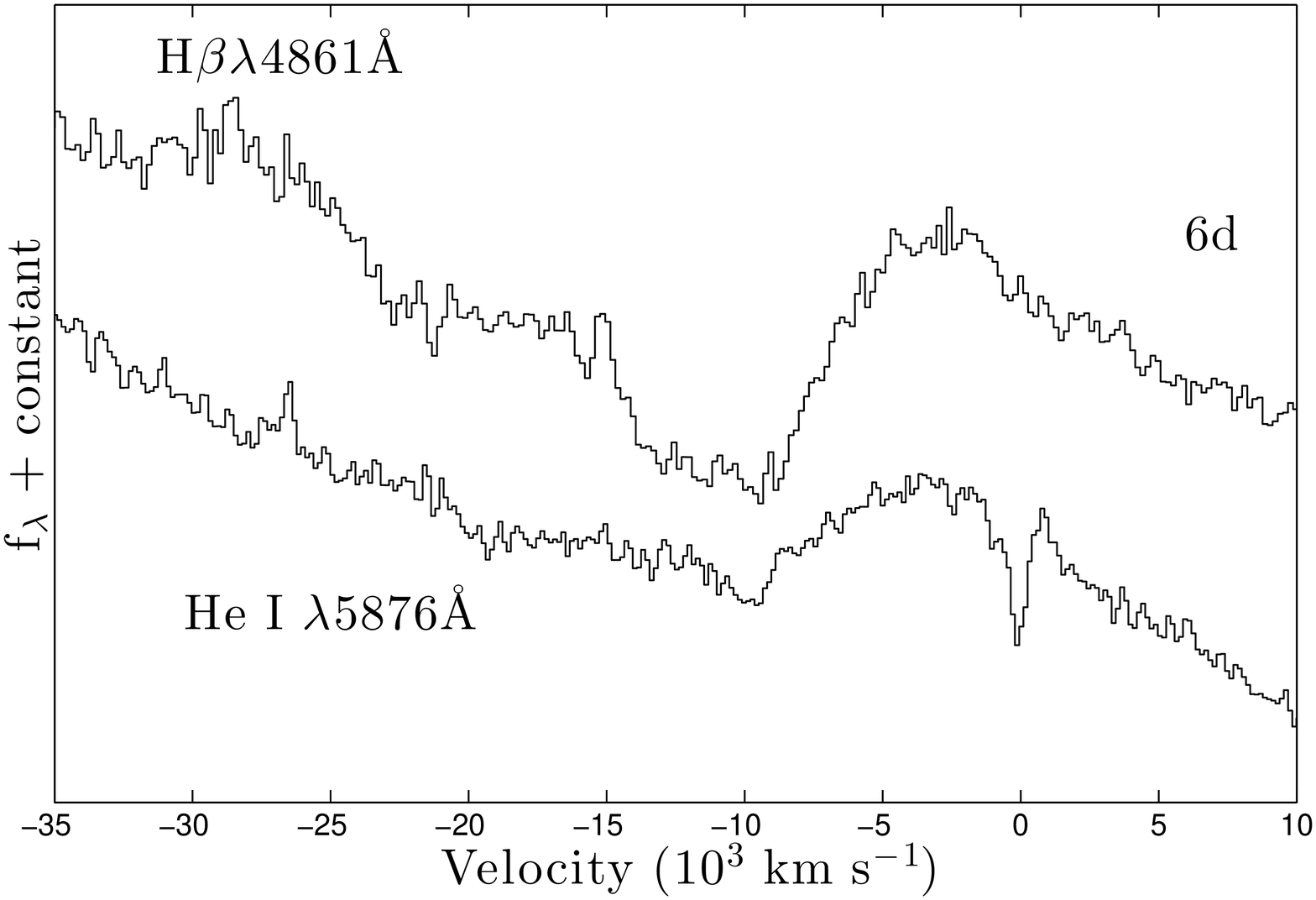}
\caption{Early-time HV features of SN\,2001X (top) and SN\,1999em (bottom) near H$\beta$ and He~I $\lambda$5876. The H$\beta$ line presents a quite prominent HV absorption line, which is missing from the He~I line.}
\label{sHV2001x_early} 
\end{figure}

HV features are also present during early phases near the H$\beta$ and He~I $\lambda$5876 profiles of SN\,1999gi \citep{Leonard:2002b}, and in the H$\beta$ absorption of SN\,1999em \citep{Baron:2000, Leonard:2002a} and SN\,2001X. Figures \ref{sHV1999gi_early}--\ref{sHV2001x_early} show the early-time spectra of SNe\,1999gi, 1999em, and 2001X around those lines. The line near 5800~\AA\ is most likely a product of He~I and not Na~I~D, especially if non-LTE effects are responsible for these lines \citep{Baron:2000}. If these features are indeed associated with hydrogen and He~I $\lambda$5876, they appear at much higher velocities than those measured in later epochs. In Figures \ref{sHV1999gi_early}--\ref{sHV2001x_early} both the H$\beta$ and the He~I $\lambda$5876 profiles of SN\,1999gi on days 6--7 and day 9 in SN\,2001X reveal what seems to be a second P-Cygni profile, containing both an emission peak and an absorption dip, at $\sim -$25,000~km~s$^{-1}$. Alternatively, \citet{Dessart:2008} identify the HV feature in SN\,1999gi as He~II $\lambda$4686.

There have been several attempts to explain the origin of these strange features. Using SYNOW, \citet{Baron:2000} identified them as HV absorption of hydrogen in the spectra of SN\,1999em. They further raised the hypothesis that these are ``complicated P-Cygni profiles,'' created by non-LTE effects which alter the Balmer level populations in the mid-velocity range, and create two line-forming regions of hydrogen and helium.

\citet{Chugai:2007}, on the other hand, suggest that interaction of the ejecta with a typical red supergiant wind leads to the formation of those features. They develop a model for the ionisation and excitation of H and He in the unshocked ejecta, considering time-dependent effects and the irradiation of ejecta by X-rays in order to study the signs of circumstellar (CS) interaction. 

They base their diagnosis of CS interaction on spectroscopic observations of H$\alpha$ and He~I $\lambda$10,830 at the photospheric phase of SN\,1999em and SN\,2004dj. In their model, the interaction of the SN ejecta and the circumstellar material (CSM) results in the formation of forward and reverse shocks, which emit X-rays and ionise and excite the outer recombined layers of the unshocked ejecta. This interaction is too weak to be revealed through specific emission lines in SNe~II-P, but sufficient to be detected as HV absorption features in the blue wing of H$\alpha$ and He~I $\lambda$10,830. 
This process causes the emergence of HV absorption in the blue wing of H$\alpha$ at $t \approx 40$--80~d and another component in He~I $\lambda$10,830 at $\sim 20$--40~d, both at a radial velocity of $\sim 10^{4}$~km~s$^{-1}$. The depth of these features increases with wind density. \citet{Chugai:2007} measured a much lower velocity for SN\,2004dj than SN\,1999em (8200 vs. 11,500~km~s$^{-1}$). Based on this, they ruled out the possibility that the line was an unidentified metal line. Similar differences in velocities at comparable epochs can be seen in Figure \ref{f:HV1}.

\citet{Chugai:2007} further suggest an additional mechanism to account for the narrow notch that appears in mid-plateau spectra of SN\,1999em and SN\,2004dj. They raise the possibility that a cool, dense shell might form at the interface of the SN and the CSM through radiative cooling. When the shell is excited by X-rays it can produce narrow HV absorption components in H$\alpha$. This produces two components: a narrow notch, created in the weakly perturbed shell, and a broad notch originating in the area of Rayleigh-Taylor instability occurring in a decelerating shell (the instability increases the turbulent velocity and broadens the line). Those components are distinguished from the effect of the ionised unshocked ejecta discussed previously, which produces a broader shallow absorption. We detect the narrow notch in the H$\alpha$ profile of SN\,1999em around day 50 and also in spectra of SN\,2001cm (day 81 -- earliest spectrum), SN\,2001X  (day 91, in both features), and SN\,2003hl (day 73). This interpretation explains the evolution of the HV feature in SN\,1999em described above.

\section{Conclusions}
Our photometric and spectroscopic analysis of a sample of 23 SNe~II-P reveals the following.

\begin{itemize}
\item The luminosity distribution of SNe~II-P seems continuous, spanning 3~mag, as indicated by \citet{Arcavi:2012}. 

\item Plateau durations are typically 100 days, as shown by \citet{Arcavi:2012}, but do get shorter for more luminous, energetic SNe, as expected from \citet{Poznanski:2013}. 

\item We do not find any indication for a correlation between rise time and plateau luminosity using a sample of 8~SNe, contrary to what was previously suggested  by \citet{GalYam:2011} and \citet{Valenti:2013}. 

\item Three SNe show a post-plateau rebrightening, which is interpreted as being caused by helium recombination by \citet{Chieffi:2003}. 

\item We find that hydrogen and iron velocities follow well-behaved power laws, with little scatter, as shown by \citet{Nugent:2006}, though we derive a somewhat steeper decline using a much larger sample. 

\item Signs of interaction with CSM might be evident in the spectra of at least six events, where we find HV features in the blue wing of H$\alpha$ during the plateau phase. Corresponding absorption in H$\beta$ confirms that these are indeed HV lines associated with hydrogen and not metal lines. 

\item When studying SNe there are a multitude of methods used in the literature to correct for dust extinction: photometric methods based on comparison to low-extinction SNe, or spectroscopic ones using the Na~I~D doublet. With the availability of a sample we were given the opportunity to ask a simple question: Does the uniformity of the sample increase after the application of a given method? Any reasonably behaved underlying distribution should become tighter after correction. The uniform answer was negative; no method we tested made a significant improvement. This is likely due to a combination of the weakness of the methods (i.e., the intrinsic scatter they introduce) and  modest typical extinction of SNe~II, of order $\sim 0.2$~mag. 

\end{itemize}

\section*{Acknowledgments}
We thank D. Maoz and I. Arcavi for helpful comments on this manuscript. 
A. Barth, 
A. Coil, 
E. Gates, 
B. Swift, 
and D. Wong 
 participated in the many observations that made this work possible, and we thank them for it. Some of the data presented herein were obtained at the W.~M. Keck Observatory, which is operated as a scientific partnership among the California Institute of Technology, the University of California and the National Aeronautics and Space Administration; it was made possible by the generous financial support of the W.~M. Keck Foundation. The authors wish to recognise and acknowledge the very significant cultural role and reverence that the summit of Mauna Kea has always had within the indigenous Hawaiian community.  We are most fortunate to have the opportunity to conduct observations from this mountain. The Kast spectrograph on the Shane 3-m reflector at Lick Observatory resulted from a generous donation made by Bill and Marina Kast. We thank the dedicated staffs of the Lick and Keck Observatories for their assistance. This research made use of the Weizmann interactive supernova data repository  (\texttt{www.weizmann.ac.il/astrophysics/wiserep}), as well as the NASA/IPAC Extragalactic Database (NED) which is operated by the Jet Propulsion Laboratory, California Institute of Technology, under contract with NASA. 

KAIT (at Lick Observatory) and its ongoing operation were made possible by donations from Sun Microsystems, Inc., the Hewlett-Packard Company, AutoScope Corporation, Lick Observatory, the NSF, the University of California, the Sylvia \& Jim Katzman Foundation, and the TABASGO Foundation. D.P. acknowledges support from the Alon fellowship for outstanding young researchers, and the Raymond and Beverly Sackler Chair for young scientists. D.C.L. acknowledges support from NSF grants AST-1009571 and  AST-1210311. J.M.S. is supported by an NSF Astronomy and Astrophysics Postdoctoral Fellowship under award AST-1302771. A.V.F.'s group at UC Berkeley has received generous financial assistance from the Christopher R. Redlich Fund, the Richard and Rhoda Goldman Fund, the TABASGO Foundation, and the NSF (most recently through grants AST-0908886 and AST-1211916).

\begin{table*}
\caption{Partial Photometric Data of the KAIT Sample.\label{t:phot}}
\begin{tabular}{ccccccc}
\hline\hline
SN name &
MJD &
Age \tablenotemark{a} &
$B$($\sigma_B$) &
$V$($\sigma_V$) &
$R$($\sigma_R$) &
$I$($\sigma_I$)\\
\hline
1999em	  & 51481.39	 & 5.4	& 13.704(0.010)	 & 13.782(0.010)	 & 13.621(0.010)	 & 13.640(0.035) \\ 
1999em	  & 51482.43	 & 6.5	& 13.677(0.010)	 & 13.718(0.010)	 & 13.577(0.010)	 & 13.578(0.036) \\ 
1999em	  & 51483.43	 & 7.5	& 13.675(0.010)	 & 13.713(0.010)	 & 13.539(0.010)	 & 13.568(0.031) \\ 
...	  & ...	 & ...	 & ...	 & ...	 & ...	 \\ 
1999gi	  & 51523.52	 & 5.4	& --	 & 14.736(0.020)	 & 14.552(0.011)	 & 14.381(0.020) \\ 
1999gi	  & 51524.56	 & 6.4	& 14.861(0.034)	 & 14.744(0.033)	 & 14.481(0.012)	 & 14.286(0.013) \\ 
1999gi	  & 51527.48	 & 9.3	& 14.763(0.034)	 & 14.614(0.013)	 & 14.299(0.010)	 & 14.116(0.010) \\ 
...	  & ...	 & ...	 & ...	 & ...	 & ...	 \\ 
2001x	  & 51975.55	 & 12.9	& 15.161(0.014)	 & 15.169(0.010)	 & 14.973(0.012)	 & 14.974(0.032) \\ 
2001x	  & 51983.52	 & 20.9	& 15.371(0.012)	 & 15.178(0.010)	 & 14.915(0.011)	 & 14.875(0.018) \\ 
2001x	  & 51988.50	 & 25.8	& 15.681(0.065)	 & 15.257(0.072)	 & --	 & -- \\ 
...	  & ...	 & ...	 & ...	 & ...	 & ...	 \\ 
2004du	  & 53230.35	 & 2.5	& 16.882(0.020)	 & 17.069(0.019)	 & 16.994(0.023)	 & 17.056(0.051) \\ 
2004du	  & 53231.30	 & 3.5	& 16.828(0.024)	 & 16.954(0.025)	 & 16.856(0.017)	 & 16.948(0.041) \\ 
2004du	  & 53233.26	 & 5.5	& 16.767(0.019)	 & 16.780(0.019)	 & 16.671(0.030)	 & 16.533(0.035) \\ 
...	  & ...	 & ...	 & ...	 & ...	 & ...	 \\ 
\hline
\end{tabular}
\tablenotetext{a}{Days since explosion}
\end{table*}

\begin{table*}
\caption{Sample Journal of Spectroscopic Observations}
\begin{tabular}{cccccc}
\hline\hline
SN Name &
MJD &
Age (Days) \tablenotemark{a} &
Instrument &
Range (\AA)&
Exposure (s)\\
\hline
2001x	 & 51971.54	 &8.9	 &Kast	 & 3300-10400 &900 \\ 
2001x	 & 51989.44	 &26.8	 &Kast	 & 3300-7820 &600 \\ 
2001x	 & 51997.00	 &34.3	 &LRIS	 & 4350-6870 &4400 \\ 
2001x	 & 51998.52	 &35.9	 &Kast	 & 3300-10400 &600 \\ 
2001x	 & 52015.00	 &52.3	 &LRIS	 & 4350-6870 &5200 \\ 
2001x	 & 52029.38	 &66.7	 &Kast	 & 3300-7840 &900 \\ 
2001x	 & 52045.36	 &82.7	 &Kast	 & 3300-7800 &900 \\ 
2001x	 & 52054.00	 &91.3	 &Kast	 & 4290-7060 &2100 \\ 
2001x	 & 52089.39	 &126.7	 &Kast	 & 3300-10400 &600 \\ 
2001x	 & 52106.34	 &143.7	 &Kast	 & 3300-10400 &1000 \\ 
2001x	 & 52144.21	 &181.6	 &Kast	 & 3300-10400 &1800 \\ 
...	 & ...	 &...	 &...	 & ...	 &... \\ 
\hline
\end{tabular}
\label{t:spec}
\tablenotetext{a}{Days since explosion}
\end{table*}

\bibliography{MYbib}
\bibliographystyle{mn2e} 
\end{document}